\documentclass{article}

\usepackage{algorithm}
\usepackage[margin=30mm,includehead,includefoot]{geometry}
\usepackage[noend]{algpseudocode}
\usepackage{amsmath,amssymb,amsthm,amscd,color,comment}
\usepackage{autobreak}
\usepackage{color}
\usepackage{xcolor}
\usepackage{url}
\usepackage{blkarray}
\usepackage{jheppub}
\usepackage{empheq}
\usepackage{graphicx}
\usepackage{booktabs, multirow}
\usepackage{subfig}

\theoremstyle{plain}

\theoremstyle{definition}
\newtheorem{example}{Example}

\newtheorem{definition}{Definition}

\usepackage{xpatch}
\makeatletter
\AtBeginDocument{\xpatchcmd{\@thm}{\thm@headpunct{.}}{\thm@headpunct{}}{}{}}
\makeatother

\newcommand{\brk}[1]{(#1)}
\newcommand{\lrbrk}[1]{\left(#1\right)}
\newcommand{\bigbrk}[1]{\bigl(#1\bigr)}
\newcommand{\Bigbrk}[1]{\Bigl(#1\Bigr)}

\newcommand{\Bigsbrk}[1]{\Bigl[#1\Bigr]}

\newcommand{\brc}[1]{\{#1\}}

\newcommand{\bigbrc}[1]{\bigl\{#1\bigr\}}

\newcommand{\abs}[1]{|#1|}
\newcommand{\vev}[1]{\langle #1\rangle}

\newcommand{\supbbf}[1]{^{\brk{\mathbf{#1}}}}
\newcommand{\supbf}[1]{^{\mathbf{#1}}}
\newcommand{\subbf}[1]{_{\mathbf{#1}}}

\newcommand{\dd}{\mathrm{d}}
\newcommand{\dlog}{\dd \log}

\newcommand{\defas}{:=}

\newcommand{\twist}{u}

\newcommand{\Integers}{\mathbb{Z}}

\newcommand{\Complex}{\mathbb{C}}
\newcommand{\CP}{\mathbb{C}\mathrm{P}}

\newcommand{\poles}{\mathbb{P}_{\omega}}
\newcommand{\Poles}{\mathbb{P}}
\newcommand{\zeros}{\mathbb{Z}_{\omega}}
\newcommand{\vv}[1]{{\boldsymbol{#1}}}
\renewcommand{\Re}{\mathrm{Re}}
\newcommand{\im}{\mathrm{i}}
\newcommand{\supp}{\mathrm{supp}}

\newcommand{\cB}{\mathcal{B}}

\newcommand{\cS}{\mathcal{S}}

\makeatletter
    \newcommand*\bigcdot{\mathpalette\bigcdot@{1}}
    \newcommand*\smtimes{\mathpalette\smtimes@{.7}}
    \newcommand*\bigcdot@[2]{\mathbin{\vcenter{\hbox{\scalebox{#2}{$\m@th#1\bullet$}}}}}
    \newcommand*\smtimes@[2]{\mathbin{\vcenter{\hbox{\scalebox{#2}{$\m@th#1\times$}}}}}
\makeatother

\newcommand{\soft}[1]{\textsc{#1}}

\newcommand{\code}[1]{\texttt{#1}}

\newcommand{\namedref}[2]{\hyperref[#2]{#1~\ref*{#2}}}
\newcommand{\secref}[1]{\namedref{Section}{#1}}

\newcommand{\appref}[1]{\namedref{Appendix}{#1}}

\newcommand{\tabref}[1]{\namedref{Table}{#1}}

\newcommand{\figref}[1]{\namedref{Figure}{#1}}

\newcommand{\exref}[1]{\namedref{Example}{#1}}

\makeatletter
\def\mr@ignsp#1 {\ifx\:#1\@empty\else #1\expandafter\mr@ignsp\fi}%
\newcommand{\multiref}[1]{\begingroup
\xdef\mr@no@sparg{\expandafter\mr@ignsp#1 \: }%
\def\mr@comma{}%
\@for\mr@refs:=\mr@no@sparg\do{\mr@comma\def\mr@comma{,\,}\ref{\mr@refs}}%
\endgroup}
\makeatother
\renewcommand{\eqref}[1]{(\multiref{#1})}

\newcommand{\be}{\begin{equation}}
\newcommand{\ee}{\end{equation}}
\newcommand{\bea}{\begin{eqnarray}}
\newcommand{\eea}{\end{eqnarray}}
\newcommand{\bei}{\begin{itemize}}
\newcommand{\eei}{\end{itemize}}

\newcommand{\jj}{\varphi}
\newcommand{\phiL}{\jj_L}
\newcommand{\phiR}{\jj_R}

\newcommand{\Res}{{\rm Res}}

\makeatletter
\newlength{\apb@width}
\newcommand{\autoparbox}[2][c]{\settowidth{\apb@width}{#2}\parbox[#1]{\apb@width}{#2}}
\newcommand{\includegraphicsbox}[2][]{\autoparbox{\includegraphics[#1]{#2}}}
\makeatother

\definecolor{green1}{HTML}{244819}
\definecolor{cyan1}{HTML}{37cdaa}
\definecolor{blue1}{HTML}{5d7ac4}
\definecolor{red1}{HTML}{d0482a}
\definecolor{purple1}{HTML}{845ea8}
\definecolor{orange1}{HTML}{e07229}

\def\done#1{ }

\title{Intersection Numbers from Higher-order Partial Differential Equations}

\author[a,b]{Vsevolod Chestnov,}
\author[c]{Hjalte Frellesvig,}
\author[a,b,d]{Federico Gasparotto,}
\author[b]{Manoj K. Mandal,}
\author[a,b]{\\Pierpaolo Mastrolia}

\newcommand{\unipd}{Dipartimento di Fisica e Astronomia, Universit\`a degli Studi di Padova,
Via Marzolo 8, I-35131 Padova, Italy.}

\newcommand{\pdinfn}{INFN, Sezione di Padova,
Via Marzolo 8, I-35131 Padova, Italy.}

\affiliation[a]{\unipd}
\affiliation[b]{\pdinfn}
\affiliation[c]{Niels Bohr International Academy, University of Copenhagen,
Blegdamsvej 17, 2100 K{\o}benhavn {\O}, Denmark}
\affiliation[d]{Johannes Gutenberg-Universit{\"a}t Mainz, D-55099 Mainz, Germany}

\emailAdd{vsevolod.chestnov@pd.infn.it}
\emailAdd{hjalte.frellesvig@nbi.ku.dk}
\emailAdd{fgasparo@uni-mainz.de}
\emailAdd{manojkumar.mandal@pd.infn.it}
\emailAdd{pierpaolo.mastrolia@unipd.it}

\abstract{We propose a new method for the evaluation of
intersection numbers for twisted meromorphic $n$-forms, through Stokes' theorem in $n$ dimensions.
It is based on the solution of an $n$-th order partial differential equation and on the evaluation of multivariate residues.
We also present an algebraic expression for the contribution from each multivariate residue.
We illustrate our approach with a number of simple examples from mathematics and physics.
}

\begin{document}
\maketitle

\section{Introduction}
\label{sec:intro}

The intersection number of differential $n$-forms
\cite{matsumoto1994,matsumoto1998,OST2003,doi:10.1142/S0129167X13500948,goto2015,goto2015b,Yoshiaki-GOTO2015203,Mizera:2017rqa,matsubaraheo2019algorithm, Ohara98intersectionnumbers, https://doi.org/10.48550/arxiv.2006.07848, https://doi.org/10.48550/arxiv.2008.03176, https://doi.org/10.48550/arxiv.2104.12584}
acts like an inner
product in the de Rham twisted cohomology group, namely the quotient space of the
closed forms {\it modulo} the exact forms, and therefore, it yields the
decomposition of differential forms in a basis of forms, generating the vector
space \cite{Mastrolia:2018uzb,Frellesvig:2019kgj,Frellesvig:2019uqt, Frellesvig:2020qot}.

A wide class of integral functions, such as Aomoto-Gel'fand integrals,
Euler-Mellin integrals, Gel’fand-Kapranov-Zelevinsky integrals, to name a few,
which embed and generalise Feynman integrals, can be considered as the pairing
of regulated integration domains, known as twisted cycles, and of regulated
forms, known as twisted cocycles
\cite{aomoto2011theory}.
Within this definition, the
integrand appears as the product of a multivalued function, called {\it
twist}, and of a differential form. The twist carries information on the
integral regularisation: for the case of dimensionally regulated Feynman
integrals, the space-time dimensionality appears in the exponent of the twist. In
this fashion, the algebraic properties of the integrals can be thought as
coming more fundamentally from the algebraic properties of the
corresponding cycles and cocycles. In particular, evaluation of
intersection numbers for twisted differential forms becomes a crucial operation
to derive linear and quadratic relations for integrals mentioned above, and to
systematically derive differential and difference equations the latter obey
\cite{Mastrolia:2018uzb,Frellesvig:2019kgj,Frellesvig:2019uqt, Frellesvig:2020qot, Mizera:2019vvs, Chen:2020uyk, Weinzierl:2020gda,Caron-Huot:2021xqj,Caron-Huot:2021iev,Chen:2022lzr}.
See ~\cite{Mizera:2019ose,Cacciatori:2021nli,Abreu:2022mfk,Weinzierl:2022eaz,Mastrolia:2022tww,Frellesvig:2021vem,Mandal:2022vok},
for recent reviews, and \cite{Chen:2022lzr,Ma:2021cxg} for applications to multi-loop calculus.
Applications of intersection theory and co-homology to diagrammatic coaction
have been presented
in~\cite{Abreu:2019wzk,Abreu:2019xep,Abreu:2021vhb,Abreu:2022erj}, and to other
interesting physical contexts in
~\cite{Mizera:2017cqs,Kaderli:2019dny,Kalyanapuram:2020vil,Weinzierl:2020nhw}.

The evaluation of the intersection numbers for twisted forms is based on the
twisted version of Stokes' theorem \cite{cho1995}. In particular, for the case
of logarithmic $n$-forms, intersection numbers can be computed by applying the
algorithm proposed in \cite{matsumoto1998} or by means of the global residue
theorem \cite{Mizera:2017rqa}. For generic meromorphic $n$-forms, the
evaluation procedure can become computationally more demanding, and it can be
performed by means of an {\it iterative approach}, as proposed in
\cite{Mizera:2019gea,Frellesvig:2019uqt,Frellesvig:2020qot}, elaborating on
\cite{Ohara98intersectionnumbers}. The iterative approach has been further
refined in \cite{Weinzierl:2020xyy}, by exploiting the invariance of the
intersection numbers for forms belonging to the same cohomology classes. In
\cite{Caron-Huot:2021xqj,Caron-Huot:2021iev}, this algorithm has been extended
to account also for the relative cohomology cases \cite{matsumoto2018relative},
required when dealing with singularities of the integrand which are not
regulated by the twist.

As an alternative to the evaluation procedure based on the Stokes' theorem,
intersection numbers can also be computed by solving the {\it secondary equation}
built from the {\it Pfaffian systems}
\cite{Matsubara-Heo-Takayama-2020b,matsubaraheo2019algorithm,Chestnov:2022alh}.
Within this algorithm, the determination of Pfaffian systems obeyed by the
generators of the cohomology group is required, and efficient methods for their
derivation have recently started to be proposed by means of Macaulay matrix in
\cite{Chestnov:2022alh}.

In this article, we propose a new algorithm for the computation of the
intersection number of twisted $n$-forms, based on a novel way of applying
Stokes' theorem that requires the solution of a {\it higher-order partial
differential equation} and application of the {\it multivariate residues}.
The computational algorithm hereby proposed can be considered as a natural
extension of the univariate case \cite{cho1995}, and, just as the latter, its
application requires the solution of a (partial) differential equation around
each intersection point, that belongs to the set of zeroes of the twist.
In this work, we show that the solution of the differential equation can be
found analytically by multiple Laurent series expansions, and that each
residue admits a closed expression in terms of the Laurent coefficients of
the two forms entering the pairing and of the twist.

The structure of the paper is as follows:
in \secref{sec:intersectionnumbers} we discuss aspects of twisted cohomology
theory and intersection numbers; we introduce a new method for 
computation of multivariate intersection numbers as multivariate residues
using a higher-order partial differential equation, and discus its
solution locally around each intersection point.
\secref{sec:examples} contains application of our
new approach to integrals and functions of interests for physics and mathematics.
In \secref{sec:algebraic} we give a closed, algebraic expression for each
residue, contributing to the multivariate intersection number.
\secref{sec:conclusion} contains our concluding remarks.
The paper includes four appendices:
\appref{app:matsumoto_v2} contains the link of our new approach to
Matsumoto's algorithm, explicitly shown in the simple case of 2-forms;
\appref{app:details} contains further details of the examples discussed in
\secref{sec:examples};
\appref{app:proof} contains the derivation of the algebraic expression given in
\secref{sec:algebraic}.

\section{Intersection numbers for twisted {\bf n}-forms}
\label{sec:intersectionnumbers}

\subsection{Twisted cohomology}
\label{sec:twisted}
Let $\cB_i$, with $i=1,\ldots,m$ , be complex homogeneous polynomials in the
homogeneous coordinates $Z=\brk{Z_1, \ldots, Z_{n + 1}}$ of the complex
projective space $\CP^n$. We introduce a manifold
$X=\CP^{n} - \bigcup_{i=1}^m {\cS}_i$,
where the hypersurfaces ${\cS}_i$ are identified by
the equations:
\begin{align}
    \cS_i \defas \bigbrc{Z \> | \> \cB_i\brk{Z} = 0}
    \ .
    \label{eq:S-def}
\end{align}
In the following we work in the chart $Z_1 \neq 0$ with the local
coordinates \brk{see~\appref{app:proj} for details}
\begin{align}
    z
    = \brk{z_1, \ldots, z_n}
    \defas \brk{{Z_2}/{Z_1}, \ldots, {Z_{n + 1}}/{Z_1}}
    \ .
\end{align}
We introduce the Aomoto-Gel'fand {\it integrals}, defined as {\it twisted period integrals},
\begin{align}
    \int_{\Gamma\supbbf{n}} u \> \varphi\supbbf{n}
    \equiv
    \int_{\Gamma\supbbf{n}} u \> \hat{\varphi}\supbbf{n} \>
    \dd^{\bf n} z \ , \qquad {\rm with} \quad
    \dd^{\bf n} z :=
    \dd z_1 \wedge \ldots \wedge \dd z_n
    \ ,
    \label{eq:twisted-int-def}
\end{align}
where:
$u$ is a multivalued function called the {\it twist}, which regulates the
integral; $\Gamma\supbbf{n}$ is a {\it regularised cycle} called {\it twisted or
loaded cycle}, {\it i.e.} an $n$-chain with empty boundary on $X$ \brk{usually
$\Gamma\supbbf{n}$ is denoted as $\Gamma\supbbf{n} \equiv \Gamma\supbbf{n}
\otimes u$ to separate the integration domain $\Gamma\supbbf{n}$ and
a specific choice of the branch of multivalued $u$ along it};
$\varphi\supbbf{n}$ is a meromorphic differential $n$-form defined
on $X$, called the {\it twisted cocycle}.
In general $u$ is a multivalued function that
``vanishes'' on the integration boundary: $u\brk{\partial{\Gamma\supbbf{n}}}=0$.
The latter property ensures that for any generic $({\bf n-1})$-form
$\varphi\supbbf{n - 1}$ the integral of the total differential is zero:
\bea
0 = \int_{\Gamma\supbbf{n}} \dd (u \, \varphi\supbbf{n - 1}) = \int_{\Gamma\supbbf{n}} u \, \nabla_\omega \, \varphi\supbbf{n - 1} \ ,
\eea
where we introduced the {\it covariant derivative}:
\bea
    \nabla_\omega \defas \dd + \omega \wedge = u^{-1} \cdot \dd \cdot u \ , \qquad {\rm with } \qquad
    \omega \equiv \sum_{i = 1}^n \hat{\omega}_i \> \dd z_i = \dlog(u) \ ,
    \label{eq:cov-deriv}
\eea
with
$
\dd = \sum_{i=i}^n \dd_{z_i} \ ,
$
where $
\dd_{z_i} = \partial_{z_i} \dd z_i \ ,
$
and
$\hat{\omega}_i = \twist^{-1} \partial_{z_i} \twist$\ ,
using the shorthand notation $\partial_{z_i} \equiv \partial/\partial z_i$\ .
When dealing with individual integration variables, it might be convenient to
consider the decomposition of the full covariant derivative:
\bea
    \nabla_\omega = \sum_{i=1}^n \nabla_{\omega_i} \ ,
\eea
with the partial covariant derivatives defined as:
\bea
    \nabla_{\omega_i} := {\hat \nabla}_{\omega_i} \dd z_i \ ,
    \qquad {\rm and } \quad
    {\hat \nabla}_{\omega_i} = \partial_{z_i} + {\hat \omega}_i \ .
\eea

Aomoto-Gel'fand integrals represent a wide class of special functions,
such as Gau\ss{} hypergeometric functions, Lauricella functions, and their
generalizations, Euler-type integrals, and Feynman integrals
\cite{Mastrolia:2018uzb}.
Integrals of this class are invariant under shifting of the differential
$n$-form by a covariant derivative: $\varphi\supbbf{n} \to \varphi\supbbf{n} +
\nabla_\omega \, \varphi\supbbf{n - 1}$\ ,
explicitly:
\bea
    \int_{\Gamma\supbbf{n}} u \, \varphi\supbbf{n} =
    \int_{\Gamma\supbbf{n}} u \, (\varphi\supbbf{n} + \nabla_\omega \, \varphi\supbbf{n - 1})
    \ .
    \label{eq:int-inv}
\eea
Similar results are obtained also for the so called {\it dual integrals},
obtained from the integrals~\eqref{eq:twisted-int-def} by replacing
$u \to u^{-1}$ and $\omega \to -\omega$ in definition~\eqref{eq:cov-deriv}.

In the case of Feynman integrals in the Baikov representation~\cite{Baikov:1996iu}, the
twist $\twist$ admits the following factorized form:
\begin{align}
    u = \prod_{i=1}^m {\cB}_i^{\gamma_i}
    \ ,
    \label{eq:B-def}
\end{align}
where the exponents $\gamma_i$ are non-integer, and the factors $\cB_i$
are polynomials build out of the kinematical and Baikov variables \brk{corresponding to
propagators}.
For this set of functions, analyticity, unitarity, and algebraic structure
are related to the geometry captured by the {\it Morse function}
$h \defas \Re(\log(u))$, see \cite{Witten:2010cx, Lee:2013hzt}.

The multivalued twist $u$ carries information on the {\it regularization}: for
dimensionally regulated Feynman integrals, it depends on the integration
variables as well as on the external scales, such as the Mandelstam invariants and
masses (all appearing in the polynomials $\cB_i$), and on the space-time
dimensionality $d$ (appearing in the $\gamma_i$). The topological information of integrals and
dual integrals is contained in $\omega$ that is a differential form with {\it
zeroes} and {\it poles}\footnote{
    We view $\omega$ as a meromorphic form on the complex projective space
    $\CP^n$, so it might also have singularities at ``infinity'',
    see~\secref{sec:examples} for some explicit examples and
    \appref{app:proj} for details on projective geometry.
}, collected in the respective sets:
\begin{align}
    \zeros = \bigbrc{\text{zeros of $\omega$}}
    \ , \quad \text{and} \quad
    \poles = \bigbrc{\text{poles of $\omega$}}
    \ .
\end{align}

The invariance of integrals and dual integrals under the
transformation~\eqref{eq:int-inv} can be used to expose the algebraic
structure of Aomoto-Gel'fand integrals.
Let us introduce two vector spaces of twisted cocycles:
the {\it twisted} $n$-th {\it cohomology group},
\begin{align}
H\supbf{n}_\omega &= \frac{{\rm Ker}(\nabla_\omega: \varphi\supbbf{n} \to \varphi\supbbf{n + 1} )}{{\rm Im}(\nabla_\omega: \varphi\supbbf{n - 1} \to \varphi\supbbf{n}) } \ ,
\label{eq:cohom-def}
\end{align}
which is the quotient space of closed $n$-forms
$\brc{\varphi\supbbf{n} \, | \ \nabla_\omega \varphi\supbbf{n} = 0}$
modulo exact forms
$\brc{\varphi\supbbf{n} \, | \ \varphi\supbbf{n} = \nabla_\omega
{\varphi\supbbf{n - 1}}}$;
and the dual twisted cohomology group:
$(H\supbf{n}_\omega)^{\vee} \defas H\supbf{n}_{-\omega}$\ .
These spaces are isomorphic, and their dimension:
\begin{equation}
    \nu = {\rm dim}(H\subbf{n}^{\pm\omega}) \ ,
\end{equation}
can be determined by counting the number of {\it critical points} of ${\cB_i}$\ ,
namely $\nu = {\rm dim}({\mathbb Z}_{\omega})$ \cite{Lee:2013hzt},
or equivalently from the Euler characteristics $\chi( {\mathbb P}_{\omega}) $ of the projective variety generated by the poles of $\omega$, as $\nu = (-1)^n (n + 1 - \chi( {\mathbb P}_{\omega} ))$ \cite{Frellesvig:2019uqt}, see also \cite{Bitoun:2017nre, Agostini:2022cgv}, or
by the Shape Lemma \cite{Frellesvig:2020qot}.

We denote the elements of the twisted cohomology~\eqref{eq:cohom-def} as
 $\langle \varphi_L| \in H\supbf{n}_\omega $~,
 $| \varphi_R \rangle \in H\supbf{n}_{-\omega} $~,
and use them to define the following natural {\it twisted Poincar{\'e} pairings}:
\begin{itemize}
    \item Integrals:
        \begin{equation}
            I
            = \int_{\Gamma\supbbf{n}} u \, \varphi_L \ .
        \end{equation}
    \item Dual integrals:
        \begin{equation}
            {I^\vee}
            = \int_{\Gamma\supbbf{n}} u^{-1} \, \varphi_R \ .
        \end{equation}
    \item Intersection numbers for twisted cocycles:
        \begin{equation}
            \langle \varphi_L \>|\> \varphi_R \rangle
            \equiv
            \brk{2 \pi \im}^{-n}
            \int_X (u \, \varphi_{L,c} ) \wedge (u^{-1} \, \varphi_R)
            \ ,
            \label{eq:interx-def}
        \end{equation}
        where $\varphi_{L,c}$ is a compactly supported cocycle equivalent to $\varphi_L$ \cite{matsumoto1998}
        \brk{see also~\secref{sec:multivardifeq} below and~\appref{app:matsumoto_v2}}.
\end{itemize}

\subsection{Integrals and Relations}
\label{ssec:integrals}

Let us now briefly review some practical applications of the twisted cohomology
theory to the study of Feynman integrals, focusing on the IBP decomposition
method.

Consider the bases generating the cohomology groups introduced in
eq.~\eqref{eq:cohom-def}:
$\{\langle e_i|\}_{i=1,\ldots,\nu} \in H^n_\omega$ and
$\{|h_i \rangle\}_{i=1,\ldots,\nu} \in H^n_{-\omega}$\ .
These two bases can be used to express the identity operator in the cohomology
space \cite{Mastrolia:2018uzb,Frellesvig:2019kgj} as follows:
\begin{eqnarray}
\mathbb{I}_c =
\sum_{i,j=1}^{\nu} | h_i \rangle \left( \mathbf{C}^{-1} \right)_{ij} \langle e_j |
\ ,
\label{eq:identity}
\end{eqnarray}
where we defined the \emph{metric matrix}:
\begin{equation}
    \label{eq:metric_matrix}
    \mathbf{C}_{ij} \defas \langle e_i \>|\> h_j \rangle \ ,
\end{equation}
whose elements are {\it intersection numbers} of the twisted basis forms.

Linear relations for Aomoto-Gel'fand-Feynman integrals, the differential
equations, and the finite difference equation they obey are consequences of
the purely algebraic application of the identity operator~\eqref{eq:identity},
see also \cite{Mastrolia:2018uzb}.

In the context of Feynman integral calculus, the decomposition of scattering
amplitudes in terms of master integrals (MIs), as well as the equations obeyed by the latter, are
derived by means of IBPs \cite{Tkachov:1981wb, Chetyrkin:1981qh} and of the Laporta method
\cite{Laporta:2001dd}.
These relations emerge naturally from the algebraic properties of twisted
cocycles.

Indeed, generic twisted cocycles can be projected onto the bases in the
corresponding vector spaces as:
\begin{eqnarray}
    \label{eq:masterdeco:cohomology}
    \langle \varphi_L| = \langle \varphi_L| \> \mathbb{I}_c  = \sum_{i=1}^{\nu} c_i \, \langle e_i| \ ,
    \quad\;\;\; & {\rm with} & \qquad
    c_i = \sum_{j=1}^{\nu} \, \langle \varphi_L \>|\> h_j \rangle \, \left(
        \mathbf{C}^{-1} \right)_{ji} \ .
\end{eqnarray}
The latter formula is dubbed the \emph{master decomposition formula} for
twisted cocycles \cite{Mastrolia:2018uzb,Frellesvig:2019kgj}. It
implies that the decomposition of any Aomoto-Gel'fand-Feynman integral
as a linear combination of MIs is an algebraic operation
(similarly to the decomposition/projection of any vector within a
vector space), which can be carried out by computing intersection
numbers of the twisted de Rham differential forms.
Using the master decomposition formula, a Feynman integral $I$ can be
decomposed in terms the MIs $J_i$ as:
\begin{eqnarray}
    I = \sum_{i=1}^{\nu} c_i \, J_i \ .
    \label{eq:decomposition_XIs}
\end{eqnarray}
with the decomposition coefficients $c_i$ given by
eq.~\eqref{eq:masterdeco:cohomology}.

Let us remark, that the metric matrix~\eqref{eq:metric_matrix}, in
general, differs from the identity matrix. The Gram-Schmidt algorithm can be
employed to build orthonormal bases from generic sets of independent elements,
using the intersection numbers as scalar products. But more generally the
coefficients appearing in the formulas~\eqref{eq:masterdeco:cohomology,
eq:decomposition_XIs} are independent of the respective dual elements.
Therefore, exploiting this freedom in choosing the corresponding dual bases may
yield striking simplifications
\cite{Frellesvig:2019kgj,Caron-Huot:2021xqj,Caron-Huot:2021iev}.
The decomposition formulas hold also in the case of the {\it
relative} twisted de Rham cohomology, which allows for relaxation of the non-integer
condition for the exponents $\gamma_i$ that appear in eq.~\eqref{eq:B-def},
see~\cite{matsumoto2018relative,Caron-Huot:2021xqj,Caron-Huot:2021iev}.

\subsection{Partial Differential Equation}
\label{sec:multivardifeq}

By elaborating on the method proposed in \cite{matsumoto1998}, we hereby
propose to evaluate the intersection number for {\bf n}-forms, using the
multivariate Stokes' theorem, yielding \brk{see also \appref{app:matsumoto_v2}}:
\bea
    \label{eq:manyvarsinterX}
    \langle \jj_L^{({\bf n})} \>|\> \jj_R^{({\bf n})}\rangle
    = ( 2 \pi \im)^{-n} \,  \int_X
    (u \, \jj_{L,c}^{({\bf n})}) \wedge (u^{-1} \jj_R^{({\bf n})})
    =
    \sum_{p \in \Poles_\omega} \Res_{z=p} (\psi \, \jj_R^{({\bf n})}) \ ,
\eea
where:
\begin{itemize}
    \item
        $\psi$ is a function (0-form), that obeys the following
        {\it $n$-th order partial differential equation} ($n$PDE):
        \begin{align}
         \label{MMdeq_v1}
            \frac{\partial^{n}}{\partial z_1 \, \partial z_{2} \ldots \partial {z_n}} (u \, \psi) &=  u \, \hat{\jj}_L^{({\bf n})} \ .
        \end{align}
    \item
        $p = \brk{p_1, p_2, \ldots, p_n} \in \Poles_\omega$ is a pole of $\omega$, i.e. an
        intersection point of singular hypersurface $\cS_i$ defined in
        eq.~\eqref{eq:S-def}, at finite location or at infinity.
    \item The residue symbol stands for:
        \begin{align}
            \Res_{z = p}\brk{f}
            = \Res_{z_n={p_n}} \ldots \Res_{z_1={p_1}}(f)
            = \brk{ 2 \pi \im}^{-n} \int_{\circlearrowleft_1 \wedge \ldots \wedge \circlearrowleft_n} f \> \dd z_1 \wedge \ldots \wedge \dd z_n
            \ ,
            \label{eq:res-def}
        \end{align}
        where the integral goes over a product of small circles
        $\circlearrowleft_i$\ , each encircling the corresponding pole~$z_i = p_i$
        in the $z_i$-plane, see \cite{Griffiths1994-oh}.
\end{itemize}

Representation~\eqref{eq:manyvarsinterX} can be derived by rewriting the intersection number
integral as a flux of a certain local form $\eta$:
\bea
    \int_X
    (\twist \, \jj_{L,c}^{({\bf n})}) \wedge (u^{-1} \jj_R^{({\bf n})})
    =
    \sum_{p \in {\mathbb P}_\omega}
    \int_{D_p}  \dd_{z_1} \ldots \dd_{z_n} \eta
    \ .
    \label{eq:interx-eta}
\eea
Working term-by-term in the sum on the RHS, let us temporarily denote by
$\brk{z_1, \ldots, z_n}$ the local coordinates centered at the intersection
point $p$. As the integration domain we take the polydisc $
    D_p = \bigbrc{
        \brk{z_1, \ldots, z_n} \> \big| \> |z_1|, \dots , |z_n| \le \epsilon
    }
$
and define:
\begin{align}
    \eta
    \defas \bar{h}_1 \ldots \bar{h}_n \>
    \bigbrk{\twist \, \psi}
    \>
    \bigbrk{\twist^{-1} \jj_R\supbbf{n}}
    \ ,
    \label{eq:eta-def}
\end{align}
where $\bar{h}_i \defas 1 - h_i$ and $h_i$ is the Heaviside step-function:
\begin{align}
    {h}_i \equiv {h}\brk{z_i}
    \defas
    \begin{cases}
        1 & \text{for $\abs{z_i} < \epsilon$\ ,}
        \\
        0 & \text{otherwise,}
    \end{cases}
\end{align}
so that the differential $\dd h_i$ is localized on the circle $\abs{z_i} = \epsilon$~.
The action of the partial derivatives in eq.~\eqref{eq:interx-eta} gives:
\begin{align}
    \label{eq:dEta}
    \dd_{z_1} \ldots \dd_{z_n} \eta
    = \Bigbrk{
        \bar{h}_1 \ldots \bar{h}_n \>
        \bigbrk{\twist \, \nabla_{\omega_1} \ldots \nabla_{\omega_n} \psi}
        + \ldots +
        \brk{-1}^n
        \bigbrk{\twist \, \psi} \>
        \dd h_1 \wedge \ldots \wedge \dd h_n
    }
    \wedge
    \bigbrk{\twist^{-1} \jj_R\supbbf{n}}
    \ .
\end{align}
By requiring that the auxilary $0$-form $\psi$ is the solution of the following $n$PDE:
\bea
    \twist \nabla_{\omega_1} \ldots \nabla_{\omega_n} \psi
    =
    \twist \, \jj_L \ ,
    \label{eq:vieq}
\eea
we obtain:
\bea
    \dd_{z_1} \ldots \dd_{z_n} \eta =
    ( u \, \jj_{L,c} ) \wedge (u^{-1} \, \jj_R)
    \ ,
\eea
where the compactly supported $n$-form
$\jj_{L,c}$ is defined as:
\bea
    \label{eq:def:phiLc}
    \jj_{L,c} \defas
    \bar{h}_1 \ldots \bar{h}_n \>
    \jj_L
    + \ldots +
    \brk{-1}^n \,
    \psi \> \dd h_1 \wedge \ldots \wedge \dd h_n
    \equiv
    \nabla_{\omega_1} \ldots \nabla_{\omega_n} \bigbrk{
        \bar{h}_1 \ldots \bar{h}_n \psi
    }
    \ .
\eea
The middle expression here is equivalent to the $\jj_{L,c}$
introduced by Matsumoto in~\cite{matsumoto1998} and, therefore,
the integration of eq.~\eqref{eq:interx-eta} can be carried out
via iterated residues.
Indeed, since $\jj_R$ is a holomorphic $n$-form, in eq.~\eqref{eq:dEta}
only the last term gives a non vanishing contribution:
\bea
    \int_X
    \bigbrk{\twist \, \jj_{L,c}\supbbf{n}} \wedge
    \bigbrk{u^{-1} \jj_R\supbbf{n}}
    &=&
    \brk{-1}^n
    \sum_{p \in \poles}
    \int_{D_p}
    \bigbrk{\twist \, \psi} \>
    \dd h_1 \wedge \ldots \wedge \dd h_n
    \wedge
    \bigbrk{\twist^{-1} \jj_R\supbbf{n}}
    \nonumber \\
    &=&
    \sum_{p \in \poles}
    \int_{\circlearrowleft_1 \wedge \ldots \wedge \circlearrowleft_n}
    \psi \, \jj_R\supbbf{n}
    \nonumber \\
    &=&
    \brk{2 \pi \im}^n
    \sum_{p \in {\mathbb P}_\omega}
    \Res_{z=p}
    (
    \psi \, \jj_R\supbbf{n}
    )
    \ ,
\eea
where the product of small circles $\circlearrowleft_1 \wedge \ldots \wedge \circlearrowleft_n$
\brk{i.e. an $n$-dimensional torus}
is the distinguished boundary of the polydisc $D_p$\ . The last equation
above\footnote{
    In the derivation, we used
    $
        \int_D \dd \bar{h} \wedge f\brk{z} \> \dd z
        = \int_{\circlearrowleft} f\brk{z} \> \dd z
    $, with $ {\circlearrowleft} := \partial D $,
    to localize the integral on the boundary.
} reproduces the result shown in eq.~\eqref{eq:manyvarsinterX}.
For more details we refer the interested reader to
the discussion in \appref{app:matsumoto_v2}.

Finally, let us once again highlight the crucial eq.~\eqref{eq:vieq} and write
it as:
\bea
    \boxed{
        \nabla_{\omega_1} \nabla_{\omega_2} \ldots \nabla_{\omega_{n}} \psi
        = \jj_L\supbbf{n}
    }\ .
    \label{eq:mostimportant}
\eea
This $n$PDE, equivalent to eq.~\eqref{MMdeq_v1}, is the natural
extension of the equation $\nabla_{\omega_1} \psi = \jj_L\supbbf{1}$
presented in \cite{matsumoto1998} for the single variable case.
Equation~\eqref{eq:mostimportant} constitutes the first main result of this
communication, as it offers a new algorithm for the direct determination of the
scalar function $\psi$, hence a simpler strategy for the evaluation of the
intersection numbers between twisted $n$-forms.

\subsection{Solution}
\label{sec:solution}

The solution of eqs.~\eqref{MMdeq_v1,eq:mostimportant} can be formally written as\footnote{
In ref.~\cite{matsumoto2018relative},
the solution $\psi$ for the twisted case with regulated pole is written by considering a modified integration contour, accounting for the contribution of monodromy. It can be shown that, around each (regulated) singular point, it is equivalent to the one considered here.
}:
\bea
\label{eq:MMdeqsol}
\psi = u^{-1} \int_{z_0}^z u \, \jj_L^{({\bf n})} \ .
\eea
A crucial observation is in order. The {\it global} solution $\psi$ is, in
general, a transcendental function. The evaluation of the intersection number
in \eqref{eq:manyvarsinterX}, though, because of the calculation of residues,
requires the knowledge of $\psi$ only {\it locally} around each of the
contributing pole, say $\psi_p = \psi|_{z \to p}$\ .
To determine the {\it local} expression of $\psi$ around the point $p$, we propose two equivalent strategies:
\begin{enumerate}
    \item solution {\bf by ansatz}\footnote{
            Here and in the following, we adopt the multivariate exponent notation:
            $\brk{z - p}^\vv{a} \equiv \brk{z_1 - p_1}^{a_1} \ldots \brk{z_n - p_n}^{a_n}$\ .
        }:
        \bea
        \psi_p \equiv \psi|_{z \to p}
        =
        \sum_{\vv{a}=\vv{a}_{\rm min}}^{\vv{a}_{\rm max}}
        \psi_{p, \vv{a}} \, (z-p)^\vv{a}  \ ,
        \label{eq:ansatz}
        \eea
        whose coefficients $\psi_{p, \vv{a}}$
        can be determined by requiring the fulfillment of
        \eqref{eq:mostimportant} around the pole $p$.
        The values of the $\vv{a}_{\rm min}$ and the $\vv{a}_{\rm max}$ depend on the
        Laurent expansion of $\jj_L^{({\bf n})}$ and $\jj_R^{({\bf n})}$ around
        $p$. The determination of the Laurent coefficients $\psi_{p, \vv{a}}$
        can be carried out by solving the {\it triangular systems} of linear
        equations that arises after inserting the ansatz in the multivariate
        differential equation.
    \item solution {\bf by series expansion}:
        \bea
            \psi_p \equiv \psi\Big|_{z \to p} =
            \bigbrk{\twist^{-1}}\Big|_{z \to p}
            \int^z_p
            \bigbrk{\twist \, \phiL\supbbf{n}}\Big|_{z \to p}
            \ ,
            \label{eq:sol-int-series}
        \eea
        which can be directly obtained by a series expansion under the integral
        sign, and a subsequent multifold integration.
        Taking $p = 0$ without loss of generality, we observe that as $z \to 0$, the
        twist~\eqref{eq:B-def} admits a factorized expansion
        $u|_{z \to 0} = z^{\gamma'} \cdot \sum_{\vv{i} \ge \vv{0}} u_{\vv{i}}
        z^{\vv{i}}$
        with non-integer exponents $\gamma'$. A similar expansion
        holds for the cocycles:
        $
            \hat{\jj}_L|_{z \to 0} =
            \sum_{\vv{i} \ge \vv{\mu_L}} \jj_{L, \vv{i}} \> z^{\vv{i}}
            $
            and
            $
            \hat{\jj}_R|_{z \to 0} =
            \sum_{\vv{i} \ge \vv{\mu_R}} \jj_{R, \vv{i}} \> z^{\vv{i}}
            $
        with integer leading exponents $\vv{\mu_L}, \vv{\mu_R} \in \Integers^n$. Using
        these expansions integration in eq.~\eqref{eq:sol-int-series} can be
        done term by term via $\int^z \dd x \> x^a = z^{a + 1} / \brk{a + 1}$.
\end{enumerate}

Let us remark that the second strategy allows for a straightforward
determination of the intersection numbers bypassing any linear system solving
procedure, and it constitutes the second main result of this study.

\subsection{Simplified formulas from Rescaling}
\label{sec:rescaling}

In the previous sections we saw how the intersection number was given by
\begin{align}
    \langle \varphi_L \>|\> \varphi_R \rangle = \sum_{p \in \Poles_\omega}
    \Res_{z=p}( \psi \, \varphi_R ) \ , \qquad
    \text{where formally} \qquad \psi = u^{-1} \int u \, \varphi_L
    \ .
    \label{eq:resc-int}
\end{align}
It is not hard to see that the above expressions are invariant under the following simultaneous rescalings:
\begin{align}
u \rightarrow u/q \,,\qquad \varphi_L \rightarrow \varphi_L \, q \,,\qquad
    \varphi_R \rightarrow \varphi_R / q  \,,\qquad \psi \rightarrow \psi \, q \,,
\end{align}
where $q$ may be any\footnote{
    To be precise, the function $q$ here has to be meromorphic and have all poles and zeroes regulated by $u$.
} function of $z$.
Since each individual residue of eq.~\eqref{eq:resc-int} possesses this invariance, the rescaling may be done locally at each individual $p$.
Such rescalings are of interest since certain specific choices for $q$ introduce simplifications.
Two special choices of $q$ are of particular interest:

\paragraph{\bf Right rescaling.}
This is defined as the choice $q = \hat{\varphi}_R$. It corresponds to the rescaling:
\begin{align}
    u \rightarrow u_R \defas u/\hat{\varphi}_R \,,\qquad
    \varphi_L \rightarrow \phi \defas \varphi_L \, \hat{\varphi}_R \,,\qquad
    \hat{\varphi}_R \rightarrow 1  \,,\qquad
    \psi \rightarrow f \defas \psi \, \hat{\varphi}_R \,,
    \label{eq:rightrescalingdef}
\end{align}

In this case, the $n$PDE eq.~\eqref{eq:mostimportant} becomes
\begin{align}
{\hat \nabla}_{\omega_{R,1}} \cdots {\hat \nabla}_{\omega_{R,n}}
f = {\hat \phi}
\,,
\label{eq:mPDRright}
\end{align}
where $\omega_R := d \log(u_R)$,
and the argument of the residues appearing in eq.~\eqref{eq:resc-int} is directly given by $f $.
Therefore,
upon the right rescaling,
it is sufficient to know the solution of the differential equation in \eqref{eq:mPDRright} up to the simple pole term,
because the higher order terms, from $\mathcal{O}(1)$ on, would not contribute to the residue.
This observation turns in a computational advantage, and,
for this reason, eq.~\eqref{eq:mPDRright},
which is a special form of eq.~\eqref{eq:mostimportant},
will become in~\secref{sec:algebraic} the starting point
for deriving an algebraic expression of ${\rm Res}(f)$.

\paragraph{\bf Left rescaling}
This form of rescaling is defined by the choice $q = 1/\hat{\varphi}_L$. It corresponds to
\begin{align}
    u \rightarrow u_L \defas u \, \hat{\varphi}_L \,,\qquad
    \hat{\varphi}_L \rightarrow 1 \,,\qquad
    \varphi_R \rightarrow \phi \defas \hat{\varphi}_L \, \varphi_R  \,,\qquad
    \psi \rightarrow g \defas \psi / \hat{\varphi}_L \,,
    \label{eq:leftrescalingdef}
\end{align}
 Upon this rescaling eq.~\eqref{eq:mostimportant} becomes
\begin{align}
\hat{\nabla}_{\omega_{L,1}} \cdots \hat{\nabla}_{\omega_{L,n}} \, g = 1
\,,
\end{align}
and we also define $\omega_L := d \log(u_L)$,
and where the argument of the residues appearing in eq.~\eqref{eq:resc-int}
becomes $g \, \hat{\varphi}_L \, \varphi_R$.
This equation is conceptually simpler to solve than the case with a generic form on the RHS.

Finally let us remark, that another source of simplifications is the shift invariance
of intersection numbers:
\begin{align}
    \vev{\jj_L \>|\> \jj_R}
    =
    \vev{\jj_L + \nabla_{\omega} \xi_L \>|\> \jj_R + \nabla_{-\omega} \xi_R}
    \ ,
\end{align}
which is valid for generic $\brk{n - 1}$-forms $\xi_L$ and $\xi_R$, but
discussion of this falls out of our scope.
Now we move on to applications of the formalism that was introduced here.

\section{Applications}
\label{sec:examples}

In this section we apply the $n$PDE method for computing intersection numbers
of twisted $2$-forms \brk{$n = 2$ variables}.
Explicit results presented below agree with the iterative method
of~\cite{Mizera:2019gea,Frellesvig:2019uqt,Frellesvig:2020qot}.

\subsection{Two-loop massless sunrise diagram}
\label{ssec:massless-sunrise}
\begin{figure}
    \centering
    \subfloat[Sunrise diagram]{
        \includegraphicsbox[scale=.12]{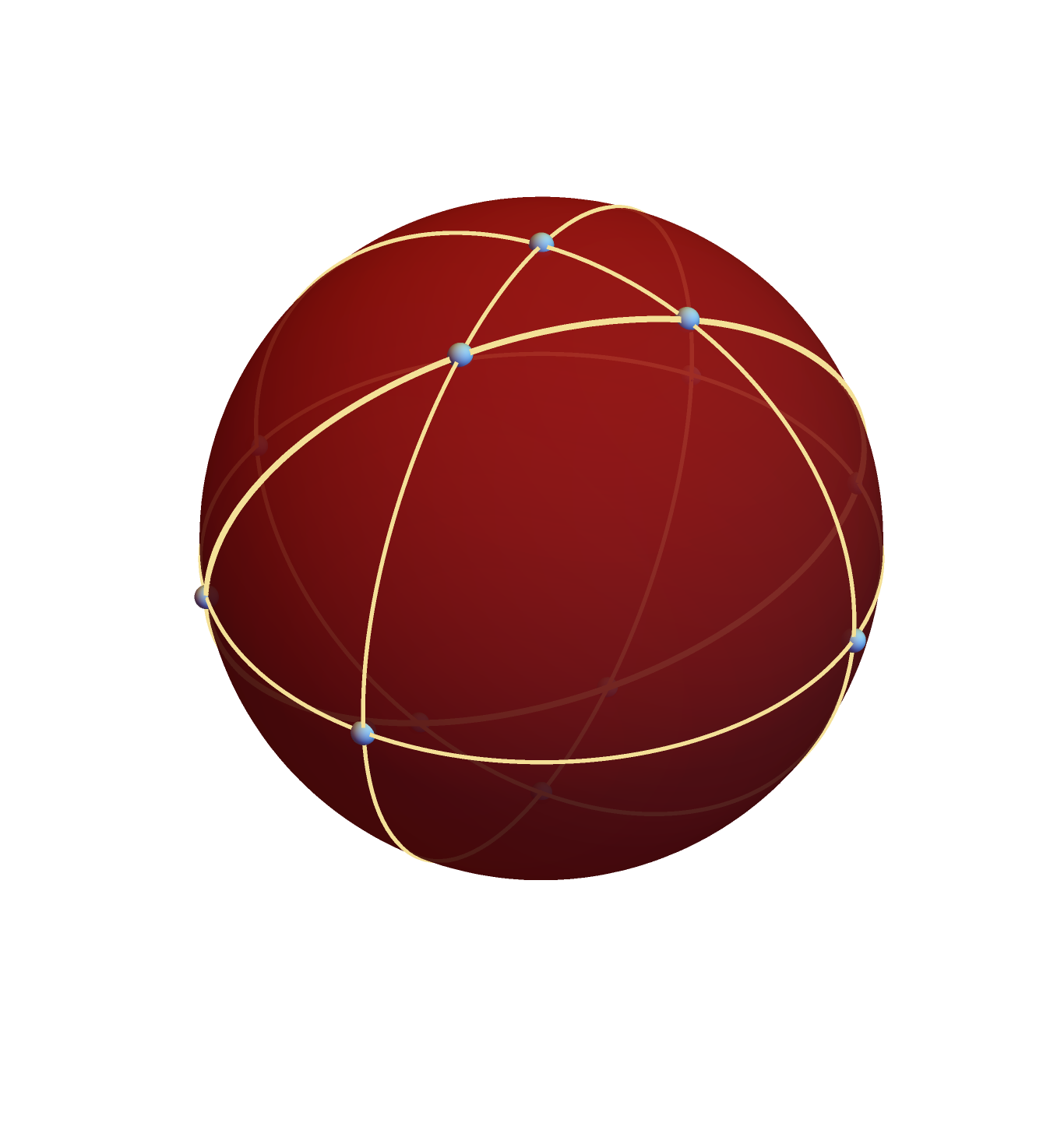}
        \label{fig:sunrise}
    }
    \subfloat[Planar box diagram]{
        \includegraphicsbox[scale=.12]{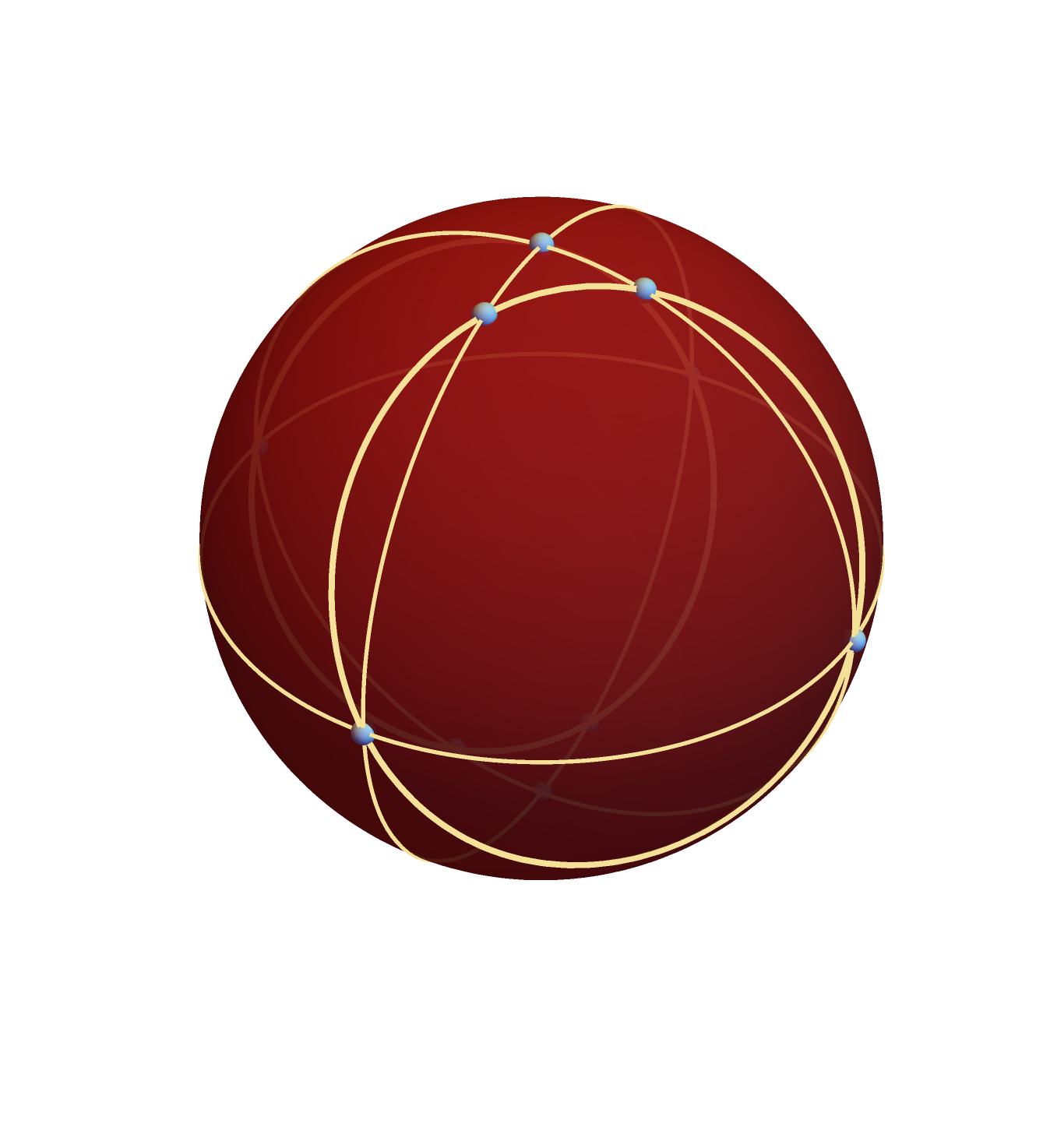}
        \label{fig:dbox}
    }
    \subfloat[${}_3F_2$]{
        \includegraphicsbox[scale=.12]{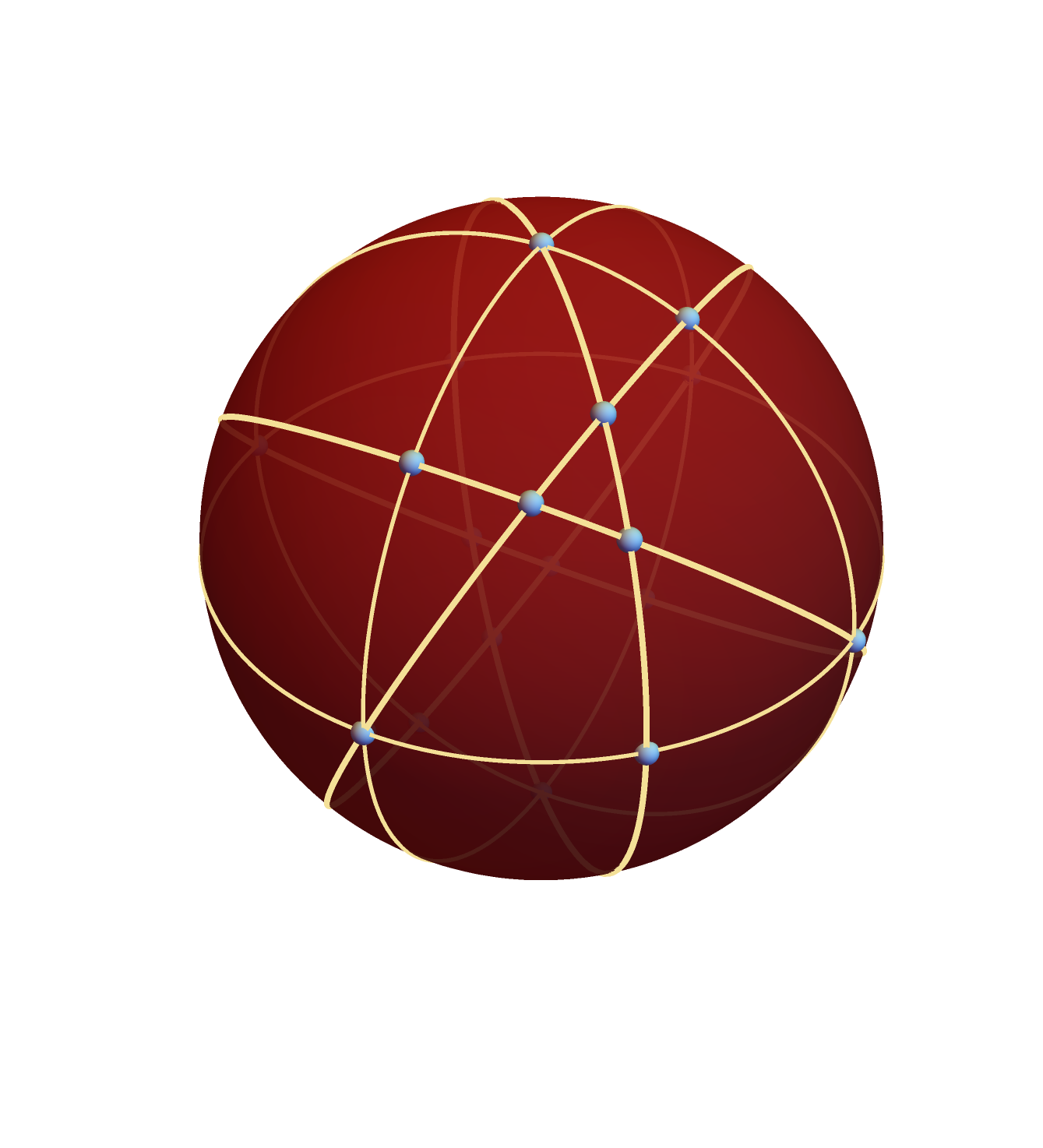}
        \label{fig:3F2}
    }
    \caption{
        Singular hypersurfaces $\cS_i$ of the massless sunrise
        twist~\protect\eqref{eq:twist-massless-sunrise} on the left,
        massless 2-loop planar box twist~\protect\eqref{eq:twist-dbox} in the middle,
        and the ${}_3F_2$ twist~\protect\eqref{eq:twist-3F2} on the right.
        The red sphere \brk{with identified antipodal points} depicts the real
        slice of the projective plane $\protect\CP^2$, whose equator is the
        line at infinity. The blue dots represent the intersection points
        collected in eqs.~\protect\eqref{eq:int-pts-massless-sunrise, eq:int-pts-dbox, eq:int-pts-3F2}.
    }
    \label{fig:sings}
\end{figure}

\begin{minipage}{12.5cm}
    The first example is the massless sunrise diagram on the maximal cut. Using the
    Baikov parametrization, we denote the two irreducible scalar products as $z_1$
    and $z_2$\,. The twist $u$ is built from the Baikov polynomial on the maximal
    cut. In the three coordinate charts $U_z$, $U_x$, and $U_y$ of the
    projective plane $\CP^2$ \brk{see~\appref{app:proj} for a brief review} it looks
    like this:
\end{minipage}
\begin{minipage}{3cm}
    \centering
    \includegraphicsbox{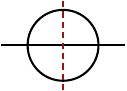}
\end{minipage}
\begin{align}
    \twist =
    \begin{cases}
        z_1^{\gamma_1} z_2^{\gamma_2}
        \>
        \brk{z_1 + z_2 - s}^{\gamma_3}
        &
        \text{in $U_z$}\ ,
        \\
        x_1^{\gamma_1} x_2^{-\gamma_0}
        \>
        \brk{1 + x_1 - s x_2}^{\gamma_3}
        &
        \text{in $U_x$}\ ,
        \\
        y_1^{-\gamma_0} y_2^{\gamma_2}
        \>
        \brk{1 - s y_1 + y_2}^{\gamma_3}
        &
        \text{in $U_y$}\ ,
    \end{cases}
    \label{eq:twist-massless-sunrise}
\end{align}
where $\gamma_0 \defas \gamma_1 + \gamma_2+ \gamma_3$\ , and the polynomial
factors~\eqref{eq:B-def} are
$\brc{\cB_1, \cB_2, \cB_3} \defas \brc{z_1, z_2, z_1 + z_2 - s}$\ .
The corresponding singular hypersurfaces $\cS_i$ defined in eq.~\eqref{eq:S-def}
intersect at $3 + 2 + 1 = 6$ points in $\CP^2$ \brk{see~\figref{fig:sunrise}}:
\begin{align}
    \poles =
    \begin{cases}
    \brk{z_1, z_2} \in \bigbrc{
        \brk{0, 0}, \brk{s, 0}, \brk{0, s}
    }
    &
    \text{in $U_z$}\ ,
    \\
    \brk{x_1, x_2} \in \bigbrc{
        \brk{0, 0}, \brk{-1, 0}
    }
    &
    \text{in $U_x$}\ ,
    \\
    \brk{y_1, y_2} = \brk{0, 0}
    &
    \text{in $U_y$}\ .
    \end{cases}
    \label{eq:int-pts-massless-sunrise}
\end{align}
In this example we consider intersection numbers between generic $\phiL\supbbf{2}$ and
$\phiR\supbbf{2}$ cocyles:
\begin{align}
    \jj\supbbf{2}_{\bigcdot} =
    z_1^{n_1} z_2^{n_2}
    \>
    \brk{z_1 + z_2 - s}^{n_3}
    \>
    \dd z_1 \wedge \dd z_2
    \ ,
    \label{eq:forms-massless-sunrise}
\end{align}
where $\bigcdot \in \brc{L, R}$\ , and $n_i \in \Integers$\ .

Our task is to compute the multivariate residues shown in
the main formula~\eqref{eq:manyvarsinterX} at each of the
points~\eqref{eq:int-pts-massless-sunrise}.
One way to do this is to find coordinate transformations that
localize on these points $\poles$\ , and allow for direct series expansion and
solution of the $n$PDE~\eqref{MMdeq_v1} either using the algebraic
formula~\eqref{eq:algebraic} or the integral formula~\eqref{eq:MMdeqsol}
\brk{see also an alternative algebraic formula in~\appref{app:closedformint}}.
We collect such transformations in~\tabref{tab:massless-sunrise-pts}
and refer the interested reader to~\appref{app:sunrise-details} for further details.

\begin{table}
    \centering
    \renewcommand{\arraystretch}{1.3}
    \begin{tabular}{cll}
        \toprule
        Chart & Point & Coordinate transformation\brk{s}
        \\\midrule
        \multirow{3}{*}{$U_z$} &
        $1: \brk{0, 0}$ &
        $\brk{z_1,\> z_2}$
        \\
        &
        $2: \brk{s, 0}$ &
        $\bigbrc{
            \brk{s + z_1 z_2,\> z_2}, \>
            \brk{s + z_1,\> z_1 z_2}, \>
            \brk{s + \brk{z_1 - 1} z_2,\> z_2}
        }$
        \\
        &
        $3: \brk{0, s}$ &
        $\bigbrc{
            \brk{z_1 z_2,\> s+z_2}, \>
            \brk{z_1,\> s + z_1 z_2}, \>
            \brk{\brk{z_1 - 1} z_2,\> s + z_2}
        }$
        \\
        \midrule
        \multirow{2}{*}{$U_x$} &
        $4: \brk{0, 0}$ &
        $\brk{\tfrac{z_1}{z_2},\> \tfrac{1}{z_2}}$
        \\
        &
        $5: \brk{-1, 0}$ &
        $\bigbrc{
            \brk{\tfrac{{z_1 z_2 - 1}}{z_2},\> \tfrac{1}{z_2}}, \>
            \brk{\tfrac{{z_1 - 1}}{{z_1 z_2}},\> \tfrac{1}{{z_1 z_2}}}, \>
            \brk{\tfrac{{\brk{s + z_1} z_2} - 1}{z_2},\> \tfrac{1}{z_2}}
        }$
        \\
        \midrule
        $U_y$ &
        $6: \brk{0, 0}$ &
        $\brk{\tfrac{1}{z_1},\> \tfrac{z_2}{z_1}}$
        \\
        \bottomrule
    \end{tabular}
    \caption{
        Intersection points $\poles$ \brk{middle} from the three coordinate
        charts of $\CP^2$ \brk{left} that contribute to the intersection
        numbers between forms~\protect\eqref{eq:forms-massless-sunrise} with
        the massless sunrise twist~\protect\eqref{eq:twist-massless-sunrise}.
        On the right we show the coordinate transformations used to
        compute the residues~\protect\eqref{eq:manyvarsinterX}.
        We sum over all the contributions of each displayed transformation.
    }
    \label{tab:massless-sunrise-pts}
\end{table}

\begin{example}
    \label{ex:sunrise-21}
    Consider the intersection $\vev{\phiL \>|\> \phiR}$ between:
    \begin{align}
        \hat{\jj}_L = \frac{1}{z_1^2 z_2} \ , \qquad
        \hat{\jj}_R = \frac{1}{z_1^2 z_2}
        \ .
    \end{align}
    Out of the six points collected in eq.~\eqref{eq:int-pts-massless-sunrise}
    only two give non-trivial residues, yielding:
    \begin{align}
        \vev{\jj_L \>|\> \jj_R} =
        \Res_{z = \brk{0, 0}}\brk{\psi \, \jj_R}
        + \Res_{z = \brk{0, s}}\brk{\psi \, \jj_R}
        =
        - \frac{1}{s^2} \frac{
            \gamma_3 \brk{\gamma_1 + \gamma_3}
        }{
            \gamma_1 \brk{1 - \gamma_1^2} \gamma_2
        }
        - \frac{1}{s^2} \frac{\gamma_3}{
            \gamma_1 \brk{1 - \gamma_1^2}
        }
        \ .
    \end{align}
\end{example}

\begin{example}
    We can also compute $\vev{\phiL \>|\> \phiR}$
    of meromorphic forms containing the $\cB_3$ factor:
    \begin{align}
        \hat{\jj}_L = \frac{1}{z_1 z_2 \brk{z_1 + z_2 - s}} \ , \qquad
        \hat{\jj}_R = \frac{1}{z_1 z_2 \brk{z_1 + z_2 - s}^2}
        \ .
    \end{align}
    Out of the six points shown in eq.~\eqref{eq:int-pts-massless-sunrise}
    the three points from the $U_z$ chart give non-zero residues:
    \begin{align}
        \Res_{z = \brk{0, 0}}\brk{\psi \, \phiR} &=
        -\tfrac{1}{s^3} \tfrac{1}{\gamma_1 \gamma_2}
        \ ,
        \\
        \Res_{z = \brk{s, 0}}\brk{\psi \, \phiR} &=
        0
        - \tfrac{1}{s^3} \tfrac{
            1 + \gamma_1 + 2 \gamma_2 + 2 \gamma_3
        }{
            \gamma_2 \brk{\gamma_2 + \gamma_3}\brk{1 + \gamma_2 + \gamma_3}
        }
        - \tfrac{1}{s^3} \Bigbrk{
            \tfrac{1 - \gamma_1}{
                \brk{1 + \gamma_3} \brk{1 + \gamma_2 + \gamma_3}
            }
            +
            \tfrac{1 + \gamma_1}{\gamma_3 \brk{\gamma_2 + \gamma_3}}
        }
        \ ,
        \\
        \Res_{z = \brk{0, s}}\brk{\psi \, \phiR} &=
        - \tfrac{1}{s^3} \tfrac{
            1 + 2 \gamma_1 + \gamma_2 + 2 \gamma_3
        }{
            \gamma_1 \brk{\gamma_1 + \gamma_3} \brk{1 + \gamma_1 + \gamma_3}
        }
        + 0
        - \tfrac{1}{s^3} \tfrac{
            \brk{1 + \gamma_1 + 2 \gamma_3}
            \brk{1 + 2 \gamma_1 + \gamma_2 + 2 \gamma_3}
        }{
            \gamma_3 \brk{1 + \gamma_3} \brk{\gamma_1 + \gamma_3}
            \brk{1 + \gamma_1 + \gamma_3}
        }
        \ ,
    \end{align}
    where we separated contributions of the transformations
    shown in~\tabref{tab:massless-sunrise-pts} into individual terms.
    By adding them up, the intersection number becomes:
    \begin{align}
        \vev{\phiL \>|\> \phiR} &=
        \Res_{z = \brk{0, 0}}\brk{\psi \, \phiR}
        + \Res_{z = \brk{s, 0}}\brk{\psi \, \phiR}
        + \Res_{z = \brk{0, s}}\brk{\psi \, \phiR}
        \nonumber \\
        &=
        -\frac{1}{s^3} \frac{
            \brk{\gamma_1 + \gamma_2 + \gamma_3} \brk{1 + \gamma_1 +
            \gamma_2 + \gamma_3}
        }{
            \gamma_1 \gamma_2 \gamma_3 \brk{1 + \gamma_3}
        }
        \ .
    \end{align}
\end{example}

\subsection{Two-loop planar box diagram}
\label{ssec:dbox}
\begin{minipage}{12.5cm}
    Now we turn to the massless 2-loop planar box on the maximal cut in the Baikov
    representation with the twist:
\end{minipage}
\begin{minipage}{3cm}
    \centering
    \includegraphicsbox{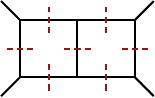}
\end{minipage}
\begin{align}
    \twist =
    \begin{cases}
        z_1^{\gamma_1} z_2^{\gamma_2} \>
        \brk{s t + s \brk{z_1 + z_2} + z_1 z_2}^{\gamma_3}
        &
        \text{in $U_z$}\ ,
        \\
        x_1^{\gamma_1} x_2^{-\gamma_0 - \gamma_3} \,
        \bigbrk{x_1 + s x_2 + s x_1 x_2 + s t x_2^2}^{\gamma_3}
        &
        \text{in $U_x$}\ ,
        \\
        y_1^{-\gamma_0 - \gamma_3} y_2^{\gamma_2} \,
        \bigbrk{s y_1 + y_2 + s y_1 y_2 + s t y_1^2}^{\gamma_3}
        &
        \text{in $U_y$}\ ,
    \end{cases}
    \label{eq:twist-dbox}
\end{align}
where $\gamma_0 \defas \gamma_1 + \gamma_2+ \gamma_3$, and the
factors~\eqref{eq:B-def} are:
\begin{align}
    \brc{\cB_1, \cB_2, \cB_3} \defas
    \brc{z_1, z_2, s t + s \brk{z_1 + z_2} + z_1}
    \ .
    \label{eq:B-dbox}
\end{align}
There are $3 + 1 + 1 = 5$ intersection points in $\CP^2$ of the singular
hypersurfaces $\cS_i$ \brk{see~\figref{fig:dbox}}, their location is as
follows:
\begin{align}
    \poles =
    \begin{cases}
    \brk{z_1, z_2} \in \bigbrc{
        \brk{0, 0}, \brk{-t, 0}, \brk{0, -t}
    }
    &
    \text{in $U_z$}\ ,
    \\
    \brk{x_1, x_2} = \brk{0, 0}
    &
    \text{in $U_x$}\ ,
    \\
    \brk{y_1, y_2} = \brk{0, 0}
    &
    \text{in $U_y$}\ .
    \end{cases}
    \label{eq:int-pts-dbox}
\end{align}
Here we only consider intersection numbers between monomial
$\phiL\supbbf{2}$ and $\phiR\supbbf{2}$ cocycles:
\begin{align}
    \jj\supbbf{2}_{\bigcdot} = z_1^{n_1} z_2^{n_2}
    \> \dd z_1 \wedge \dd z_2\ ,
    \label{eq:forms-dbox}
\end{align}
with the same notation as in eq.~\eqref{eq:forms-massless-sunrise}.

Some of the intersection points~\eqref{eq:int-pts-dbox} turn out to be
degenerate: they have three singular hypersurfaces $\cS_i$ passing
through them. An example of this is the point $x = \brk{0, 0}$ located on the
line at infinity, as can be seen from eq.~\eqref{eq:twist-dbox}
and~\figref{fig:dbox}.
One way to amend this issue is to employ the resolution of singularities
procedure \brk{closely related to the sector decomposition algorithm, see \cite{Borowka:2017idc, Heinrich:2021dbf, Bogner:2007cr}}.
In \appref{app:dbox-details} we give further details and collect the full list of
coordinate transformations used for computation of intersection numbers in this
setup.
\begin{example}
    \label{ex:dbox-11}
    Let us consider the intersection number $\vev{\phiL \>|\> \phiR}$ between the
    two logarithmic forms:
    \begin{align}
        \hat{\jj}_L = \frac{1}{z_1 z_2} \ , \qquad \hat{\jj}_R = \frac{1}{z_1 z_2}
        \ .
    \end{align}
    Only the origins of the three charts~\eqref{eq:int-pts-dbox}
    contribute to this intersection number producing:
    \begin{align}
        \vev{\phiL \>|\> \phiR}
        &= \Res_{z = \brk{0, 0}}\brk{\psi \, \phiR}
        + \Res_{x = \brk{0, 0}}\brk{\psi \, \phiR}
        + \Res_{y = \brk{0, 0}}\brk{\psi \, \phiR}
        \nonumber\\
        &=
        \tfrac{1}{\gamma_1 \gamma_2}
        + \Bigbrk{
            \tfrac{-1}{\gamma_1 \brk{\gamma_2 + \gamma_3}}
            +
            \tfrac{1}{\brk{\gamma_2 + \gamma_3} \brk{\gamma_0 + \gamma_3}}
        }
        + \Bigbrk{
            \tfrac{1}{\brk{\gamma_1 + \gamma_3} \brk{\gamma_0 + \gamma_3}}
            +
            \tfrac{-1}{\gamma_2 \brk{\gamma_1 + \gamma_3}}
        }
        \ .
    \end{align}
\end{example}

\begin{example}
    Similar to~\exref{ex:sunrise-21}, we compute the intersection between:
    \begin{align}
        \hat{\jj}_L = \frac{1}{z_1^2 z_2} \ , \qquad \hat{\jj}_R = \frac{1}{z_1^2 z_2}
        \ .
    \end{align}
    We find the three points from $\poles$ contribute to this intersection
    number, resulting in:
    \begin{align}
        \vev{\phiL \>|\> \phiR} &=
        \Res_{z = \brk{0, 0}}\brk{\psi \, \phiR}
        + \Res_{z = \brk{0, -t}}\brk{\psi \, \phiR}
        + \Res_{x = \brk{0, 0}}\brk{\psi \, \phiR}
        \nonumber\\
        &=
        - \tfrac{1}{t^2} \tfrac{\gamma_3 \brk{\gamma_1 + \gamma_3}}{
            \gamma_1 \gamma_2 \brk{1 - \gamma_1^2}
        }
        - \tfrac{1}{s^2 t^2} \tfrac{\gamma_3 \brk{s - t}^2}{
            \gamma_1 \brk{1 - \gamma_1^2}
        }
        + \tfrac{1}{s^2} \tfrac{\gamma_3 \brk{\gamma_1 + \gamma_3}}{
            \gamma_1 \brk{1 - \gamma_1^2} \brk{\gamma_2 + \gamma_3}
        }
        \ .
    \end{align}
\end{example}

\begin{table}
    \centering
    \renewcommand{\arraystretch}{1.3}
    \begin{tabular}{ccl}
        \toprule
        Chart & Point & Coordinate transformation\brk{s}
        \\\midrule
        \multirow{6}{*}{$U_z$} &
        $\brk{0, 0}$ &
        $\bigbrc{
            \brk{z_1 z_2,\> z_2},
            \brk{z_1,\> z_1 z_2},
            \brk{\brk{1 + z_1} z_2,\> z_2}
        }$
        \\
        &
        $\brk{0, s}$ &
        $\brk{z_1,\> s + z_2}$
        \\
        &
        $\brk{s, s}$ &
        $\bigbrc{
            \brk{s + z_1 z_2,\> s + z_2},
            \brk{s + z_1, \> s + z_1 z_2},
            \brk{s + \brk{1 + z_1} z_2,\> s + z_2}
        }$
        \\
        &
        $\brk{1, 0}$ &
        $\brk{1 + z_1,\> z_2}$
        \\
        &
        $\brk{1, s}$ &
        $\brk{1 + z_1,\> s + z_2}$
        \\
        &
        $\brk{1, 1}$ &
        $\bigbrc{
            \brk{1 + z_1 z_2,\> 1 + z_2},
            \brk{1 + z_1, \> 1 + z_1 z_2},
            \brk{1 + \brk{1 + z_1} z_2,\> 1 + z_2}
        }$
        \\
        \midrule
        \multirow{2}{*}{$U_x$} &
        $\brk{0, 0}$ &
        $\bigbrc{
            \brk{z_1,\> \tfrac{1}{z_2}},
            \brk{\tfrac{1}{z_2},\> \tfrac{1}{z_1 z_2}},
            \brk{1 + z_1,\> \tfrac{1}{z_2}}
        }$
        \\
        &
        $\brk{1, 0}$ &
        $\brk{\tfrac{1 + z_1}{z_2},\> \tfrac{1}{z_2}}$
        \\
        \midrule
        $U_y$ &
        $\brk{0, 0}$ &
        $\bigbrc{
            \brk{\tfrac{1}{z_1 z_2},\> \tfrac{1}{z_1}},
            \brk{\tfrac{1}{z_1},\> z_2},
            \brk{\tfrac{1}{\brk{s^{-1} + z_1} z_2}, \tfrac{1}{s^{-1} + z_1}}
        }$
        \\
        \bottomrule
    \end{tabular}
    \caption{
        Intersection points $\poles$ \brk{middle} from the three coordinate
        charts of $\CP^2$ \brk{left} that contribute to the intersection
        numbers between forms~\protect\eqref{eq:forms-3F2} with
        the ${}_3F_2$ twist~\protect\eqref{eq:twist-3F2}.
        On the right we show the coordinate transformations used to
        compute the residues~\protect\eqref{eq:manyvarsinterX}.
        We sum over all the contributions of each displayed transformation.
    }
    \label{tab:3F2-pts}
\end{table}

\subsection{${}_3F_2$ hypergeometric function}
\label{ssec:3F2}
Another setup with degenerate intersection points is the ${}_3F_2$
hypergeometric function, whose twist looks like this:
\begin{align}
    \twist =
    \begin{cases}
        z_1^{\gamma_1} z_2^{\gamma_2} \>
        \brk{1 - z_1}^{\gamma_3} \>
        \brk{s - z_2}^{\gamma_4} \>
        \brk{z_1 - z_2}^{\gamma_5}
        &
        \text{in $U_z$}\ ,
        \\
        \brk{-1}^{\gamma_3 + \gamma_5} \>
        x_1^{\gamma_1} x_2^{-\gamma_0} \>
        \brk{x_1 - x_2}^{\gamma_3} \>
        \brk{s x_2 - 1}^{\gamma_4} \>
        \brk{1 - x_1}^{\gamma_5}
        &
        \text{in $U_x$}\ ,
        \\
        \brk{-1}^{\gamma_3} \>
        y_1^{-\gamma_0} y_2^{\gamma_2} \>
        \brk{1 - y_1}^{\gamma_3} \>
        \brk{s y_1 - y_2}^{\gamma_4} \>
        \brk{1 - y_2}^{\gamma_5}
        &
        \text{in $U_y$}\ ,
    \end{cases}
    \label{eq:twist-3F2}
\end{align}
where $\gamma_0 \defas \gamma_1 + \ldots + \gamma_5$\ .
The five singular hypersurfaces $\cS_i$ intersect at the following
$6 + 2 + 1 = 9$ points \brk{see~\figref{fig:3F2}}:
\begin{align}
    \poles =
    \begin{cases}
    \brk{z_1, z_2} \in \bigbrc{
        \brk{0, 0}, \brk{0, s}, \brk{s, s}, \brk{1, 0}, \brk{1, s}, \brk{1, 1}
    }
    &
    \text{in $U_z$}\ ,
    \\
    \brk{x_1, x_2} \in \bigbrc{
        \brk{0, 0}, \brk{1, 0}
    }
    &
    \text{in $U_x$}\ ,
    \\
    \brk{y_1, y_2} = \brk{0, 0}
    &
    \text{in $U_y$}\ .
    \end{cases}
    \label{eq:int-pts-3F2}
\end{align}
We compute intersection numbers between generic $\phiL\supbbf{2}$ and
$\phiR\supbbf{2}$ cocyles:
\begin{align}
    \jj\supbbf{2}_{\bigcdot} =
    z_1^{n_1} z_2^{n_2}
    \>
    \brk{1 - z_1}^{n_3}
    \brk{s - z_2}^{n_4}
    \brk{z_1 - z_2}^{n_5}
    \>
    \dd z_1 \wedge \dd z_2
    \ ,
    \label{eq:forms-3F2}
\end{align}
with the same notation as in eq.~\eqref{eq:forms-massless-sunrise}.

\begin{example}
    In continuation of~\exref{ex:dbox-11} we
    calculate the intersection of the logarithmic forms:
    \begin{align}
        \hat{\jj}_L = \frac{1}{z_1 z_2} \ , \qquad \hat{\jj}_R = \frac{1}{z_1 z_2}
        \ .
    \end{align}
    Once again, only the origins of the charts contribute to this intersection
    number giving the total:
    \begin{align}
        \vev{\phiL \>|\> \phiR}
        &= \Res_{z = \brk{0, 0}}\brk{\psi \, \phiR}
        + \Res_{x = \brk{0, 0}}\brk{\psi \, \phiR}
        + \Res_{y = \brk{0, 0}}\brk{\psi \, \phiR}
        \nonumber\\
        &=
        \Bigbrk{
            \tfrac{1}{\gamma_1 \brk{\gamma_1 + \gamma_2 + \gamma_5}}
            +
            \tfrac{1}{\gamma_2 \brk{\gamma_1 + \gamma_2 + \gamma_5}}
            +
            0
        }
        \\
        &\quad + \Bigbrk{
            \tfrac{-1}{\gamma_1 \brk{\gamma_2 + \gamma_4 + \gamma_5}}
            +
            \tfrac{1}{\gamma_0 \brk{\gamma_2 + \gamma_4 + \gamma_5}}
            +
            0
        }
        + \Bigbrk{
            \tfrac{1}{\gamma_0 \brk{\gamma_1 + \gamma_3 + \gamma_5}}
            +
            \tfrac{-1}{\gamma_2 \brk{\gamma_1 + \gamma_3 + \gamma_5}}
            +
            0
        }
        \nonumber
        \ .
    \end{align}
\end{example}
This concludes our collection of examples, for intersection numbers in $n = 2$
variables, involving Feynman integrals and other mathematical functions.  Let
us remark, that the presented method is also applicable to meromorphic
$n$-forms in higher dimensions.

\section{Algebraic solutions}
\label{sec:algebraic}

In this section, we will introduce an algebraic solution for the contribution to the intersection number coming from each individual residue.
The main formula of this section, eq.~\eqref{eq:algebraic}, is derived by solving the $n$PDE using an ansatz. We will start the section by introducing an essential combinatorial object, that of \textit{vector compositions}.

\subsection{Vector Compositions}
\label{sec:vc}

Vector compositions are a convenient tool for manipulating
algebraic expressions involving multivariate monomials,
in particular when dealing with product of multivariate series.

\begin{definition}
Vector compositions
\begin{align}
S = \text{VC}(\boldsymbol{\tau}) \qquad \text{with} \qquad \boldsymbol{\tau} = (\tau_1,\ldots,\tau_n)
\end{align}
are a function of $n$ non-negative integers $\tau_1$ to $\tau_n$. Its output $S$
is a set of ordered lists $\sigma$, each containing a number of $n$-vectors
$\boldsymbol{s}$, with the property $\sum_{\boldsymbol{s} \in \sigma} \boldsymbol{s} = \boldsymbol{\tau}$
($\boldsymbol{s}$ vectors contain non-negative integers;
the zero-vector is excluded, corresponding to\footnote{We order integer vectors using the {\it product order}, see~\appref{app:compare} for a review.} all $\boldsymbol{s} >
\boldsymbol{0}$).
Each $\boldsymbol{s}$ can be represented as a step in  $\mathbb{Z}^n$, therefore $\sigma$, being a collection of steps, can be mapped to a path in $\mathbb{Z}^n$,
going from $\boldsymbol{0}$ to $\boldsymbol{\tau}$.
Within this representation,
$S$ can be viewed as the set of all paths
$\mathbb{Z}^n$, connecting $\boldsymbol{0}$ to $\boldsymbol{\tau}$, in one or in multiple (ordered) steps,
counted by the size of $\sigma$, namely having $|\sigma|=1,2,\ldots$ m,
with $m := \sum_{i=1}^n \tau_i \,.$
\end{definition}

\begin{example}[$n=1$ case]
    In the univariate case, the vector $\boldsymbol{\tau}$ has just one entry, say $\boldsymbol{\tau} = (m)$ with $m \in {\mathbb N}$,
    therefore the vector
    compositions reduce to standard \textit{compositions}, {\it i.e.} $\text{VC}(\boldsymbol{\tau}) = \text{comp}(m)$. Compositions are a known operation from combinatorics which describe the number of (ordered) ways an integer can be expressed as a sum of positive integers.
    For illustration, let us choose $\boldsymbol{\tau}=(4)$
    \begin{align}
        {\rm VC}(\boldsymbol{\tau}) = \text{comp}(4) = \big\{ (4), (3,1), (2,2), (1,3), (2,1,1), (1,2,1), (1,1,2), (1,1,1,1) \big\}
    \end{align}
\end{example}

\begin{example}[$n=2$ case\nopunct]
    \label{ex:n2}
    The two dimensional case, where the vector $\boldsymbol{\tau}$ has two components, is the first non-trivial case for applying vector composition. For illustration, let us consider  $\boldsymbol{\tau}=(1,2)$, yielding
    \begin{figure}[ht]
        \centering
        \includegraphicsbox[]{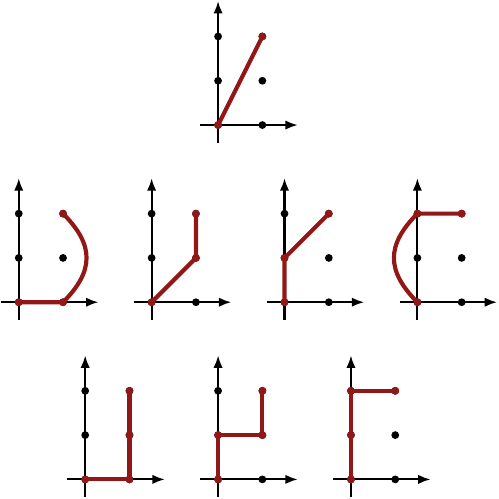}
        \caption{
            Paths appearing in eq.~\protect\eqref{eq:vcexample}.
        }
        \label{fig:paths}
    \end{figure}
    \begin{align}
        \text{VC}(\boldsymbol{\tau}) = \Big\{ & \left( \tbinom{1}{2} \right),
        \nonumber \\
        &
        \left( \tbinom{1}{0}, \tbinom{0}{2} \right),
        \left( \tbinom{1}{1}, \tbinom{0}{1} \right),
        \left( \tbinom{0}{1}, \tbinom{1}{1} \right),
        \left( \tbinom{0}{2}, \tbinom{1}{0} \right),
        \label{eq:vcexample} \\
        &\left( \tbinom{1}{0}, \tbinom{0}{1}, \tbinom{0}{1} \right), \left( \tbinom{0}{1}, \tbinom{1}{0}, \tbinom{0}{1} \right), \left( \tbinom{0}{1}, \tbinom{0}{1}, \tbinom{1}{0} \right) \Big\}
        \nonumber
    \end{align}
    where the three rows correspond to entries with $|\sigma|$ being 1, 2, and 3, respectively. Each element of
    $\text{VC}(\boldsymbol{\tau})$
    admits a representation in terms of lattice paths in $\mathbb{Z}^2$, as shown  in~\figref{fig:paths}.
\end{example}

More details
on vector compositions can be found in e.g.~\cite{andrews_1984}.
In the current context, they turn out to be a useful tool
to express the solution of the $n$PDE eq.~\eqref{eq:mPDRright}.

\subsection{Algebraic expression for residues}
\label{sec:algexp}

In~\secref{sec:intersectionnumbers} we saw that multivariate intersection
numbers can be expressed as a sum over contributions from a number of points,
with each contribution expressed as a residue. In the ``right rescaling''
framework of \secref{sec:rescaling} the relation is given as
\begin{align}
    \langle \varphi_L \>|\> \varphi_R \rangle = \sum_{p \in \Poles_\omega} \text{Res}_{z=p}(f)
    \label{eq:multivarincf}
\end{align}
where $f$ is defined as the solution of eq.~\eqref{eq:mPDRright}, namely
\begin{align}
    {\hat \nabla}_{\omega_{R,1}} \cdots
    {\hat \nabla}_{\omega_{R,n}} \, f = {\hat \phi} \ ,
    \qquad {\rm with} \;\qquad
    {\hat \phi} = \hat{\varphi}_L \hat{\varphi}_R \ ,
    \label{eq:fdef}
\end{align}
and where
\begin{align}
    {\hat \nabla}_{\omega_{R,i}} :=
    (\partial_{z_i} + (\partial_{z_i} w)) \ ,
    \qquad
    \text{with} \;\qquad w := \log(u/\hat{\varphi}_R) \ .
\end{align}
In \secref{sec:solution} it was discussed how to solve systems of the form of
eq.~\eqref{eq:fdef} by making a series ansatz for $f$ and solving it locally
around each point, in what is essentially an analytic use of the ``Frobenius method''. In this section we are going to pursue that direction a step
further, to obtain an algebraic solution for each term in
eq.~\eqref{eq:multivarincf}. The first step is to write $w$ and
$\hat{\phi}$
as series around each point, where the
appropriate variable changes have been made in order to ensure that the point
is located at $\boldsymbol{z} = \boldsymbol{0}$ and in addition that the series
expansions are well defined so no further variable changes (or blowups) are
needed. With this we may write
\begin{align}
    {\hat \phi}
    &= \sum_{\boldsymbol{i} = \boldsymbol{\mu}_{\boldsymbol{\phi}} \!\!\!}^{\boldsymbol{\infty}} \phi_{\boldsymbol{i}} z^{\boldsymbol{i}} \qquad\quad \text{and} \qquad\quad w = \log(u/\hat{\varphi}_R) = \sum_{\boldsymbol{i} = \boldsymbol{0}}^{\boldsymbol{\infty}} w_{\boldsymbol{i}} z^{\boldsymbol{i}} \; + \; \sum_{j=1}^n v_j \log(z_j)
\end{align}
where $z^{\boldsymbol{i}} := z_1^{i_1} \cdots z_n^{i_n}$. In this notation
$\boldsymbol{\mu}_{\boldsymbol{\phi}}$ should be interpreted as the powers
corresponding to the leading term in the expansion of $\hat{\varphi}_L
\hat{\varphi}_R$. The $w$-expansion begins at powers $\boldsymbol{0}$
(lower ones cannot appear
when $u$ is of the form $\prod_i \mathcal{B}_i^{\gamma_i}$, as in our case).
Using an ansatz for $f$ as a Laurent series expansion around each intersection
point, the $n$PDE constraints the coefficients, so that the residue of $f$
can be obtained in closed form as,
\begin{align}
\boxed{ \, \text{Res}(f) \, = \! \sum_{\boldsymbol{\tau} = \boldsymbol{0}}^{\! {-}\boldsymbol{\mu}_{\boldsymbol{\phi}}{-}\boldsymbol{2}} \! \sum_{\sigma \in \text{VC}(\boldsymbol{\tau})} \! \left( \frac{(-1)^{|\sigma|}}{U(\boldsymbol{0},\boldsymbol{0})} \prod_{i=1}^{|\sigma|}
\frac{U(\boldsymbol{\sigma}_{\boldsymbol{i}}, \boldsymbol{\mathcal{T}}_{\! \boldsymbol{i}} )}{U(\boldsymbol{0}, \boldsymbol{\mathcal{T}}_{\! \boldsymbol{i}} )} \right)
\! \phi_{-\boldsymbol{\tau} - \boldsymbol{2}} \,
}
\label{eq:algebraic}
\end{align}
where
\begin{align}
\boldsymbol{\mathcal{T}}_{\! \boldsymbol{i}} := \boldsymbol{\tau} - \sum_{j<i} \!
\boldsymbol{\sigma}_{\boldsymbol{j}}
\end{align}
and $U$ is a function dependent only on the terms in the expansion of $w$, which is conveniently defined in terms of an auxiliary function, $R$, as
\begin{align}
U(\boldsymbol{\lambda},\boldsymbol{\eta}) &:= R(\boldsymbol{\lambda}, \, {-}\boldsymbol{\eta}{-}\boldsymbol{1})
\end{align}
whose general definition is reported in \appref{app:Rneq2}. Here, for illustration purposes, we showcase its expressions for a few small values of $n$:
\begin{alignat}{2}
    & \bullet \ n=1:
    \quad &&R(\alpha,\beta) \,=\, (\beta{+}v) \delta_{\alpha,0} + \alpha w_{\alpha}
    \label{eq:reks1} \\[3mm]
    & \bullet \ n=2: &&R(\boldsymbol{\alpha}, \boldsymbol{\beta}) = (\beta_1{+}v_1) (\beta_2{+}v_2) \delta_{\boldsymbol{\alpha},\boldsymbol{0}}
        + \big( \alpha_1 \alpha_2 + \alpha_1 (\beta_2{+}v_2) + \alpha_2 (\beta_1{+}v_1) \big) w_{\boldsymbol{\alpha}}
    \nonumber \\
    &&&\> + \sum_{\boldsymbol{j} \,=\, \boldsymbol{0}}^{\boldsymbol{\alpha}} (\alpha_1 {-} j_1) j_2 \, w_{\boldsymbol{\alpha} - \boldsymbol{j}} w_{\boldsymbol{j}}
    \label{eq:reks2}
    \\[3mm]
    & \bullet \ n=3: \quad &&R(\boldsymbol{\alpha}, \boldsymbol{\beta}) = (\beta_1 {+} v_1) (\beta_2 {+} v_2) (\beta_3 {+} v_3) \delta_{\boldsymbol{\alpha},\boldsymbol{0}}
    \nonumber \\
    &&&\> + \big( (\alpha_1 {+} \beta_1 {+} v_1) (\alpha_2 {+} \beta_2 {+} v_2) (\alpha_3 {+} \beta_3 {+} v_3) - (\beta_1 {+} v_1) (\beta_2 {+} v_2) (\beta_3 {+} v_3) \big) w_{\boldsymbol{\alpha}}
    \nonumber \\
    &&&\> + \sum_{\boldsymbol{j} \,=\, \boldsymbol{0}}^{\boldsymbol{\alpha}} \big( (\alpha_1{-}j_1) j_2 (j_3{+}\beta_3{+}v_3) + (\alpha_2{-}j_2) j_3 (j_1{+}\beta_1{+}v_1) + (\alpha_3{-}j_3) j_1 (j_2{+}\beta_2{+}v_2) \big) w_{\boldsymbol{\alpha}{-}\boldsymbol{j}} w_{\boldsymbol{j}}
    \nonumber \\
    &&&\> + \sum_{\boldsymbol{j} = \boldsymbol{0}}^{\boldsymbol{\alpha}} \, \sum_{\boldsymbol{l} = \boldsymbol{0}}^{\boldsymbol{\alpha}{-}\boldsymbol{j}} \, (\alpha_1{-}j_1{-}l_1) \, j_2 \, l_3 \, w_{\boldsymbol{\alpha}{-}\boldsymbol{j}{-}\boldsymbol{l}} \, w_{\boldsymbol{j}} \, w_{\boldsymbol{l}} \label{eq:reks3}
\end{alignat}

\noindent
For a general $n$-variate expression of $R$, see \appref{app:Rneq2}.

The residue formula in eq.~\eqref{eq:algebraic}, using vector compositions,
constitutes the third main result of this work. Further details on its
derivation can be found in \appref{app:proof}.\\

The arXiv version of this paper is supplemented with a {\sc Mathematica} implementation of
eq.~\eqref{eq:algebraic}, consisting of a package \code{algebraic\_residue.m}
and a notebook file
\code{load\_algebraic\_residue.nb} that contains examples of
its use.

\subsection{Example}
\label{sec:algexample}

Here we show an application of the algebraic formula eq.~\eqref{eq:algebraic}.

\begin{example}
    \label{ex:sunrise-11}
    Using the setup of~\secref{ssec:massless-sunrise}, let us compute $\langle \jj_L \>|\> \jj_R \rangle$ with
\begin{align}
    \hat{\varphi}_L = \frac{1}{z_1 z_2} \,, \qquad \hat{\varphi}_R = \frac{1}{z_1 z_2}
    \ .
\end{align}
In the ``right rescaling'' scheme, the first step is to compute $\Phi = \varphi_L\hat{\varphi}_R$ at each of the points in \tabref{tab:massless-sunrise-pts} and identify the leading term corresponding to the $\boldsymbol{\mu}_{\boldsymbol{\phi}}$-vector. The result is given in \tabref{tab:muatpoints}.

\begin{table}[h!]
  \begin{center}
    \begin{tabular}{|c|c|c|}
      \hline
      \text{point} & $\hat{\Phi}$ & ${-}\boldsymbol{\mu}_{\boldsymbol{\phi}}{-}\boldsymbol{2}$ \\ \hline
      1  & $z_1^{-2} z_2^{-2}$ & $\{0,0\}$ \\ \hline
      2a & $(s{+}z_1 z_2)^{-2}$ & $\{-2,-2\}$ \\ \hline
      2b & $(s{+}z_1)^{-2} z_2^{-2}$ & $\{-2,0\}$ \\ \hline
      2c & $(s{+}z_2(z_1{-}1))^{-2} \!$ & $\{-2,-2\}$ \\ \hline
      3a & $z_1^{-2} (s{+}z_2)^{-2}$ & $\{0,-2\}$ \\ \hline
      3b & $(s{+}z_1 z_2)^{-2}$ & $\{-2,-2\}$ \\ \hline
    \end{tabular} $\;\;$ \begin{tabular}{|c|c|c|}
      \hline
      \text{point} & $\hat{\Phi}$ & ${-}\boldsymbol{\mu}_{\boldsymbol{\phi}}{-}\boldsymbol{2}$ \\ \hline
      3c & $(z_1{-}1)^{-2} (z_2{+}s)^{-2}$ & $\{-2,-2\}$ \\ \hline
      4  & $z_1^{-2} z_2^{-2}$ & $\{0,0\}$ \\ \hline
      5a & $(z_1 z_2 {-} 1)^{-2}$ & $\{-2,-2\}$ \\ \hline
      5b & $(z_1{-}1)^{-2} z_2^{-2}$ & $\{-2,0\}$ \\ \hline
      5c & $((s{+}z_1)z_2{-}1)^{-2}$ & $\{-2,-2\}$ \\ \hline
      6  & $z_1^{-2} z_2^{-2}$ & $\{0,0\}$ \\ \hline
    \end{tabular}

    \caption{Values for $\hat{\Phi}$ and $\boldsymbol{\mu}_{\boldsymbol{\phi}}$ at the twelve sub-intersection points for \exref{ex:sunrise-11}.
    The points are defined in \tabref{tab:massless-sunrise-pts}, and the additional labels (a,b,c) refer to the
    adopted coordinate transformation.
    \label{tab:muatpoints}}
  \end{center}
\end{table}

We see from the general expression~\eqref{eq:algebraic} that a given point will only contribute if $-\boldsymbol{\mu}_{\boldsymbol{\phi}}-\boldsymbol{2} \geq \boldsymbol{0}$. So we realize that there only will be contributions from three of the intersection points: 1, 4, and 6. In each case eq.~\eqref{eq:algebraic} will contain one term only, of the form
\begin{align}
\text{Res}(f) = \frac{\phi_{-2,-2}}{(v_1{-}1)(v_2{-}1)}
\ .
\end{align}
We observe that $\phi_{-2,-2} = 1$ at each of the three points, while the values of the $v_i$ depend on the considered point, and are computed locally.
The non-vanishing expressions of $\text{Res}(f)$, are:
\begin{align}
\text{Point 1} : \frac{1}{\gamma_1 \gamma_2} \qquad \text{Point 4} : \frac{-1}{\gamma_1 (\gamma_1{+}\gamma_2{+}\gamma_3)} \qquad \text{Point 6} : \frac{-1}{\gamma_2 (\gamma_1{+}\gamma_2{+}\gamma_3)}
\ ,
\end{align}
therefore, the intersection number is obtained by adding them up, as:
\begin{align}
\langle \varphi_R \>|\> \varphi_L \rangle &= \frac{\gamma_3}{\gamma_1 \gamma_2 (\gamma_1{+}\gamma_2{+}\gamma_3)}
\ ,
\end{align}
which is the correct expected results.
\end{example}

\section{Conclusion}
\label{sec:conclusion}

In this work we proposed a new computational method for the evaluation of
intersection numbers for twisted meromorphic $n$-forms through Stokes' theorem,
which is based on the solution of a $n$-th order partial differential equation
($n$PDE). Our finding can be summarised simply as:
\begin{equation}
    \langle
    \varphi_L \>|\> \varphi_R
    \rangle
    =
    \frac{1}{(2 \pi i)^n}
    \int ( u \, \varphi_{L,c} ) \wedge (u^{-1} \, \varphi_R)
    =
    \frac{1}{(2 \pi i)^n}
    \oint (u^{-1} \, \varphi_R) \int (u \, \varphi_L) \ .
\end{equation}

The evaluation of the last integral is performed by multivariate Laurent series
expansions and multivariate residues. The analytic properties of the
intersection numbers and of the $n$PDE yielded the algebraic determination of
the contributing residues, which we were able to cast in closed forms for an
efficient evaluation.

The presented method requires the knowledge of the intersection points as
input. In the case of a twist corresponding to a hyperplane arrangement and
normal crossing at the intersection points, this information can be easily
extracted from the integrand. Instead, for generic configurations, when more
hypersurfaces pass through a given intersection point, the desingularization
procedure constitutes a challenging, hence interesting, mathematical problem.

The formulation of the intersection number for twisted cohomology presented
here applies to the case of regulated singularities. We are confident that it
can be extended to the relative cohomology case, where the singularities of the
integral are not regulated within the twist.

The new method presented here can be applied to derive linear relations,
differential equations, difference equations, and quadratic relations for
Feynman integrals as well as for a wider class of functions, such as
Aomoto-Gel'fand integrals, Euler-Mellin integrals, and GKZ hypergeometric
functions. Therefore the method can be useful for computational (quantum) field
theory, and computational differential and algebraic topology.

\subsection*{Acknowledgements}

We wish to acknowledge interesting and stimulating discussions on intersection theory with Sergio Cacciatori, Yoshiaki Goto, Saiei Matsubara-Heo, Keiji Matsumoto, and Nobuki Takayama, at various stages.
We would like to thank Sebastian Mizera for discussions and comments on the manuscript, and Henrik Munch for many good discussions and various checks.
H.F. would like to thank Cristian Vergu for stimulating discussions. \\
We also would like to acknowledge the anonymous referee for her/his suggestions about improving our manuscripts.

V.C. is supported by the {\it Diagrammalgebra} Stars Wild-Card Grant UNIPD.
H.F. is partially supported by a Carlsberg Foundation Reintegration Fellowship, and has received funding from the European Union’s Horizon 2020 research and innovation program under the Marie Skłodowska-Curie grant agreement No. 847523 ‘INTERACTIONS’.
The work of M.K.M. is supported by Fellini - Fellowship for Innovation at INFN funded by the European Union's Horizon 2020 research and innovation programme under the Marie Sk{\l}odowska-Curie grant agreement No 754496.
F.G. is supported by Fondazione Cassa di Risparmio di Padova e Rovigo (CARIPARO).

\appendix

\section{Relation to Matsumoto's algorithm}
\label{app:matsumoto_v2}
In this appendix we adopt the notation of \cite{matsumoto1998} \brk{see
also~\cite{Mizera:2019gea} for a review}
and explicitly derive the $n$PDE~\eqref{eq:mostimportant} in the $n = 2$ variable setting.

Let us consider an intersection number among two \emph{holomorphic} $2$-forms $\varphi_L$ and
$\varphi_R$, and assume that any two hypersurfaces $\cS_i$ and
$\cS_j$ defined in eq.~\eqref{eq:S-def} intersect at point\brk{s}, such that no other
$\cS_k$ with $k \neq i, j$ passes through it. The main idea of
\cite{matsumoto1998} is to find an explicit $2$-form $\varphi_{L,c}$ with
compact support in the same cohomology
class of $\varphi_L$\ . This $2$-form appears in the definition
of the intersection number integral~\eqref{eq:manyvarsinterX}:
\begin{equation}
    \vev{\varphi_L \>|\> \varphi_R}
    =
    \brk{2 \pi \im}^{-2} \, \int_X \varphi_{L,c} \wedge \varphi_R\ .
    \label{eq:IN_appendix}
\end{equation}
We associate an intersection point $p_{ij} \defas \cS_i \cap \cS_j $ with each pair of
indices $\brk{i, j}$ with $i < j$.
For a singular hypersurface $\cS_i$ we introduce two tubular neighborhoods $V_i$
and $U_i$ around it of radius $\epsilon_1$ and
$\epsilon_2$ respectively, with $\epsilon_1 < \epsilon_2$\ , such that $V_i
\subset U_i$\ .
Following~\cite{matsumoto1998}, we then write an explicit formula for the
compactly supported $2$-form:
\begin{equation}
    \varphi_{L,c}
    = \varphi_L - \nabla_{\omega} \lrbrk{
        \sum_{j} h_{j} \lrbrk{
            \>\sum_{i<j} \dd h_{i} \> \psi_{p_{ij}}
            + \prod_{i<j} (1-h_{i}) \> \psi^{j}_{p_{ij}}
        }
    }\ ,
    \label{eq:iota_map_2var}
\end{equation}
where $h_{j}$ is a smoothened version of the Heaviside step function:
$h_{j}=1$ in $V_{j}$\ , $h_{j}=0$ outside $U_{j}$~, and it smoothly
interpolates between the two boundary values $0 \leq h_{j} \leq 1$ in
$U_{j} \setminus V_{j}$\ . These properties of $h_{j}$ imply that
the two forms coincide $\jj_{L,c} = \jj_L$ outside of the singular
neighborhood $\cup_j U_j \supset \cup_j \cS_j$\ .

Furthermore, it is not difficult to show that eq.~\eqref{eq:iota_map_2var}
has compact support, i.e.
the $2$-form $\jj_{L,c}$ vanishes on the singular neighborhood
$\cup_j V_j \subset \cup_j U_j$, if we impose certain constraints
on the auxiliary $0$- and $1$-forms appearing on the RHS of eq.~\eqref{eq:iota_map_2var}.
For our purposes here it is enough to investigate these constraints
locally\footnote{
    See the original work~\cite{matsumoto1998} and the review
    in~\cite{Mizera:2019gea} for the full treatment.
} around
a given intersection point $p_{ij}$ for some fixed $i$ and $j$, where
we introduce local coordinates
$\brk{z_1, z_2}$, such that the hypersurfaces $\cS_{i}$ and $\cS_{j}$ are
parametrized by $z_1=0$ and $z_2=0$ respectively.
Then the auxiliary 1-forms $\psi_{}^{i}$ and $\psi_{}^{j}$, and the 0-form
$\psi_{}$ \brk{we will drop the subscripts in the following} from
eq.~\eqref{eq:iota_map_2var} must satisfy the following
differential equations~\cite{matsumoto1998}:
\begin{alignat}{2}
    \nabla_\omega \psi_{}^{i} & = \varphi_L && \text{on $U_{i} \defas
        \bigbrc{\brk{z_1, z_2} \>\big|\> \abs{z_1} \le \epsilon_2}
    $} \ ,
    \nonumber\\
    \nabla_\omega \psi_{}^{j} & = \varphi_L && \text{on $U_{j} \defas
        \bigbrc{\brk{z_1, z_2} \>\big|\> \abs{z_2} \le \epsilon_2}
    $} \ ,
    \label{eq:set_of_DEQ_psi}
    \\
    \nabla_\omega \psi_{} & = \psi_{}^{i} - \psi_{}^{j} \qquad && \text{on $U_{i} \cap U_{j}$} \ ,
    \nonumber
\end{alignat}
with the $z$-dependence shown explicitly:
\begin{equation}
    \psi_{}^{i} = \hat{\psi}_{}^{i}\brk{z_1, z_2} \> \dd z_2\ , \qquad
    \psi_{}^{j} = \hat{\psi}_{}^{j}\brk{z_1, z_2} \> \dd z_1\ , \qquad
    \psi_{} \equiv \hat{\psi}_{}\brk{z_1, z_2}\ .
\end{equation}
Then on the polydisc $U_{i} \cap U_{j}$ eq.~\eqref{eq:iota_map_2var} reduces to:
\begin{eqnarray}
    \varphi_{L,c}
    = (1-h_{j})(1-h_{i}) \> \varphi_L
    - (1-h_{j}) \> \dd h_{i} \wedge \psi_{}^{i}
    - (1-h_{i}) \> \dd h_{j} \wedge \psi_{}^{j}
    + \psi_{} \> \dd h_{i} \wedge \dd h_{j}\ ,
    \label{eq:compact_supported_partII}
\end{eqnarray}
which vanishes on the inner polydisc $V_{i} \cap V_{j}$~. Similarly one
can show that eq.~\eqref{eq:iota_map_2var} vanishes on the rest of the singular tube
$V_{j} \setminus \cup_{i \neq j} U_{i}$ for each $j$, which proves that $\jj_{L,c}$
indeed has compact support.

In the integral~\eqref{eq:IN_appendix} the compactly supported $2$-form
$\jj_{L,c}$ is wedged against the holomorphic form $\varphi_R$, hence out of
eq.~\eqref{eq:iota_map_2var} only the anti-holomorphic part will contribute to
the intersection number.
The only source of this is the last term in eq.~\eqref{eq:compact_supported_partII},
so that the intersection number integral localizes onto
neighborhoods of intersection points $p_{ij}$ as:
\begin{equation}
    \vev{\varphi_L \>|\> \varphi_R}
    =
    \brk{2 \pi \im}^{-2}
    \sum_{p_{ij}}
    \int_{(U_{i} \setminus V_{i}) \cap (U_{j} \setminus V_{j})}
    \psi_{} \> \dd h_{i} \wedge \dd h_{j}  \wedge \varphi_R
    \ ,
    \label{eq:IN_localization}
\end{equation}
where the integration domain is a difference of two polydiscs:
\begin{equation}
    (U_{i} \setminus V_{i}) \cap (U_{j} \setminus V_{j})
    = \{ (z_1,z_2) \, | \, \epsilon_1 \leq |z_1|, |z_2| \leq \epsilon_2 \}
    \equiv \brk{U_{i} \cap U_{j}} \setminus \brk{V_{i} \cap V_{j}}
\end{equation}
Each integral in eq.~\eqref{eq:IN_localization} can be computed applying Stokes' theorem one
variable at time. Recalling that $h_{i}$ vanishes on $\partial U_{i} = \{ z_1 \, | \, |z_1|=\epsilon_2\}$
\brk{and similar for $h_{j}$},
we obtain the following:
\begin{equation}
    \begin{split}
        (2 \pi \im)^{-2}    \int_{\epsilon_1 \leq |z_1|, |z_2| \leq \epsilon_2} \dd h_{i} \wedge \dd h_{j}  \wedge \psi_{} \,  \jj_R
        & = - (2 \pi \im)^{-2} \int_{|z_1|=\epsilon_1,\epsilon_1 \leq |z_2| \leq \epsilon_2 } \dd h_{j} \wedge \psi_{} \jj_R \\
        & =-(2 \pi \im)^{-1} \int_{\epsilon_1 \leq | z_2 | \leq \epsilon_2} \dd
        h_{j} \wedge \bigbrk{ \Res_{z_1=0} \,  \psi_{} \jj_R } \\
        & = (2 \pi \im)^{-1} \int_{|z_2|=\epsilon_1} \,  \Res_{z_1=0} \,  \psi_{} \jj_R \\
        & = \Res_{z_2=0} \bigbrk{ \Res_{z_1=0} \,  \psi_{} \jj_R }
        \ .
    \end{split}
\end{equation}

We would like to emphasise, that in ref.~\cite{matsumoto1998} the
0-form $\psi_{}$ is determined via a multi-step procedure, in which the
auxiliary 1-forms $\psi_{}^{i}, \psi_{}^{j}$ have to be determined first as
solution to the system~\eqref{eq:set_of_DEQ_psi}.
In our work, instead, $\psi_{}$ is determined at one go as the solution of a
single higher-order differential equation:
\begin{equation}
    \nabla_{\omega_1} \nabla_{\omega_2} \psi_{}
    = \varphi_L
    \ , \qquad \text{on $U_{i} \cap U_{j}$}\ .
\end{equation}
In fact, considering $\nabla_{\omega}= \nabla_{\omega_1} + \nabla_{\omega_2}$,
we have the following identity:
\begin{eqnarray}
    \nabla_{\omega_2}\psi_{} = \psi_{}^{i}\ .
\end{eqnarray}
Then the last equation of eq.~\eqref{eq:set_of_DEQ_psi} can be rearranged as:
\begin{eqnarray}
    0 = \nabla_{\omega} \left( \nabla_{\omega} \psi_{} - (\psi_{}^{i} - \psi_{}^{j}) \right)= \nabla_{\omega}
    \lrbrk{\nabla_{\omega_1} \psi_{} + \psi_{}^{j}}\ ,
    \label{eq:0_nabla2}
\end{eqnarray}
which indeed gives the $n$PDE eq.~\eqref{eq:mostimportant} in $n = 2$ variables:
$\nabla_{\omega_1} \nabla_{\omega_2} \psi_{} = \varphi_L$\ .

\subsection{Examples of $\eta$}
Here for illustrative purposes we collect the two examples of the $\eta$
introduced in eq.~\eqref{eq:eta-def}.
\begin{example}[$n = 1$ case]
    \label{ex:eta1}
    Locally around a given intersection point the univariate
    version of eq.~\eqref{eq:eta-def} is:
    \begin{align}
        \eta =
        \bar{h}_1 \>
        \bigbrk{\twist \, \psi} \>
        \bigbrk{\twist^{-1} \jj_R}
        \ .
    \end{align}
    The corresponding version of the intersection number
    integral~\eqref{eq:interx-eta} reads:
    \begin{align}
        \dd_1 \eta
        &= \Bigbrk{
            \bar{h}_1 \>
            \bigbrk{\twist \nabla_{\omega_1} \psi}
            -
            \bigbrk{\twist \, \psi} \>
            \dd h_1
        }
        \wedge \bigbrk{\twist^{-1} \jj_R}
        \\
        &\equiv
        \bigbrk{\twist \jj_{L,c}}
        \wedge \bigbrk{\twist^{-1} \jj_R}
        \ ,
    \end{align}
    where the compactly supported $1$-form~\eqref{eq:def:phiLc} becomes:
    \begin{align}
        \jj_{L,c} =
        \jj_{L} - \nabla_{\omega_1} \brk{h_1 \psi}
        \ .
    \end{align}
    in agreement with~\cite{matsumoto1998}.
\end{example}
\begin{example}[$n = 2$ case]
    Specifying the formula~\eqref{eq:eta-def} to $2$-variate case gives:
    \label{ex:eta2}
    \begin{align}
        \eta =
        \bar{h}_1 \bar{h}_2 \>
        \bigbrk{\twist \, \psi} \>
        \bigbrk{\twist^{-1} \jj_R}
        \ ,
    \end{align}
    so that the integrand~\eqref{eq:interx-eta} turns into:
    \begin{align}
        \dd_1 \dd_2 \eta
        &= \Bigbrk{
            \bar{h}_1 \bar{h}_2 \>
            \bigbrk{\twist \nabla_{\omega_1} \nabla_{\omega_2} \psi}
            -
            \bar{h}_2 \>
            \dd h_1 \wedge
            \bigbrk{\twist \nabla_{\omega_2} \psi}
            \nonumber\\
            &\quad-
            \bar{h}_1 \>
            \bigbrk{\twist \nabla_{\omega_1} \psi} \wedge
            \dd h_2
            +
            \bigbrk{\twist \, \psi} \>
            \dd h_1 \wedge \dd h_2
        }
        \wedge \bigbrk{\twist^{-1} \jj_R}
        \\
        &\equiv
        \bigbrk{\twist \jj_{L,c}}
        \wedge \bigbrk{\twist^{-1} \jj_R}
        \ ,
    \end{align}
    where the compactly supported $2$-form $\jj_{L,c}$ agrees with the
    definition~\eqref{eq:compact_supported_partII}.
\end{example}

\section{Additional details for the applications}
\label{app:details}
In this appendix we provide further details for the examples presented in
\secref{sec:examples}.

\subsection{Projective plane}
\label{app:proj}
Here we review some of the basic features of the complex projective space
$\CP^n$, which is the ambient space of the integration domain $X$ introduced in
eq.~\eqref{eq:interx-def}. To support the examples discussed
in~\secref{sec:examples}, we focus on the $n = 2$ dimensional case known as
the complex projective plane $\CP^2$.

Consider a triple of complex variables $\brk{Z_1, Z_2, Z_3} \in \Complex^3
\setminus \brk{0, 0, 0}$\ . The projective plane $\CP^2$ consists of
equivalence classes of these triples under a uniform rescaling:
$\brk{Z_1, Z_2, Z_3} \equiv \brk{\lambda Z_1, \lambda Z_2, \lambda Z_3}$
for $\lambda \in \Complex^\ast$. In other words, $\CP^2$ is the space of complex
lines in $\Complex^3$ passing through the origin.

We turn $\CP^2$ into a manifold by covering it with three coordinate charts
defined as follows:
\begin{align}
    \brc{U_z, U_y, U_x}_i \defas \bigbrc{\brk{Z_1, Z_2, Z_3} \> \big| \> Z_i \ne
    0}_{i \in \brc{1, 2, 3}}\ ,
    \label{eq:charts-def}
\end{align}
that is, the sets of triples with one non-zero coordinate $Z_i$\ , which we can
safely rescale away.
Locally in these charts $\CP^2$ looks like a copy of $\Complex^2$, so we
introduce three pairs of local coordinates shown in \tabref{tab:cp2-charts}.

In the examples below we determine the distinct intersection points
of various curves $\cS_i$ \brk{see above eq. \eqref{eq:twisted-int-def}} inside of
$\CP^2$. Therefore it is important to understand how the charts overlap with
one another in order to avoid overcounting of points. This can be easily
deduced from the definition \eqref{eq:charts-def}: for example, the $U_z \cap
U_y$ is given by triples with $Z_1 \neq 0$ and $Z_2 \neq 0$, while $U_y
\setminus U_z$ is determined by $Z_1 = 0$ and $Z_2 \neq 0$.

Thus in the following we will work mainly in the $U_z$ chart along with the corresponding line at
infinity: the projective line $\CP^1$ given by $Z_1 = 0$\ . This line
passes through the other two charts $U_x$ and $U_y$~, where it is determined
by $x_2 = 0$ and $y_1 = 0$ respectively.

\begin{table}
    \centering
    \renewcommand{\arraystretch}{1.3}
    \begin{tabular}{ccc}
        \toprule
        Chart & Local coordinates &
        Back to $U_z$
        \\\midrule
        $U_z$ &
        $\brk{1, z_1, z_2} \defas
        \brk{1, \tfrac{Z_2}{Z_1}, \tfrac{Z_3}{Z_1}}$ &
        \\
        $U_x$ &
        $\brk{x_2, x_1, 1} \defas
        \brk{\tfrac{Z_1}{Z_3}, \tfrac{Z_2}{Z_3}, 1}$ &
        $\brk{x_1, x_2} = \brk{\tfrac{z_1}{z_2}, \tfrac{1}{z_2}}$
        \\
        $U_y$ &
        $\brk{y_1, 1, y_2} \defas
        \brk{\tfrac{Z_1}{Z_2}, 1, \tfrac{Z_3}{Z_2}}$ &
        $\brk{y_1, y_2} = \brk{\tfrac{1}{z_1}, \tfrac{z_2}{z_1}}$
        \\
        \bottomrule
    \end{tabular}
    \caption{
        The three coordinate charts \brk{left} that cover $\protect\CP^2$ with
        their local coordinates \brk{middle}.
        On the right we show the coordinate
        transformations from the $U_y$ and $U_x$ charts back to $U_z$.
        In this section we focus on the $U_z$ chart and its line at
        infinity, i.e. the projective line $\CP^1$ given by $Z_1 = 0$.
    }
    \label{tab:cp2-charts}
\end{table}

\subsection{Two-loop massless sunrise diagram}
\label{app:sunrise-details}
Here we elaborate on the material of \secref{ssec:massless-sunrise},
namely we discuss the coordinate transformations presented in
\tabref{tab:massless-sunrise-pts}, focusing on the intersection point $z =
\brk{s, 0}$ as an illustrative example.

The main purpose of our coordinate transformations is to compute the
multivariate residue in the main formula~\eqref{eq:manyvarsinterX} at the
selected intersection point.
So we start with a shift $\brk{z_1, z_2} \mapsto \brk{s + z_1, z_2}$,
after which the twist~\eqref{eq:twist-massless-sunrise} turns into:
\begin{align}
    \twist =
    \brk{s + z_1}^{\gamma_1} z_2^{\gamma_2}
    \>
    \brk{z_1 + z_2}^{\gamma_3}
    \ .
    \label{eq:twist-shift}
\end{align}
The last factor here is problematic: its series expansion around the new
origin $z_1 = z_2 = 0$ is ill-defined. For example, expanding
$1 / \brk{z_1 + z_2}$ in the following two ways we get different results:
\begin{align}
    1 / \brk{z_1 + z_2} =
    \begin{cases}
        z_2^{-1} - z_1 z_2^{-2} + \ldots
        &
        \text{first in $z_1$, then in $z_2$\ ,}
        \\
        z_1^{-1} - z_1^{-2} z_2 + \ldots
        &
        \text{first in $z_2$, then in $z_1$\ .}
    \end{cases}
\end{align}
To fix this we further change
$\brk{z_1, z_2} \mapsto \brk{z_1, z_1 z_2}$,
so that the twist~\eqref{eq:twist-shift} becomes regular:
\begin{align}
    \twist =
    z_1^{\gamma_2 + \gamma_3} z_2^{\gamma_2}
    \brk{s + z_1}^{\gamma_1}
    \>
    \brk{1 + z_2}^{\gamma_3}
    \ ,
    \label{eq:twist-shift-blowup}
\end{align}
meaning that all the polynomial factors now have constant terms. Series
expansion of eq.~\eqref{eq:twist-shift-blowup} no longer depends on the order
of operations and we may proceed with evaluation of the multivariate residue
in eq.~\eqref{eq:manyvarsinterX} as discussed in
\secref{sec:intersectionnumbers} and \secref{ssec:massless-sunrise}.
Putting together the transformations that led us to the
twist~\eqref{eq:twist-shift-blowup} gives:
\begin{align}
    \brk{z_1, z_2}
    \mapsto
    \brk{s + z_1, z_1 z_2}
    \ ,
    \label{eq:shift-blowup}
\end{align}
where we recognize one of the elements listed on the second row
of~\tabref{tab:massless-sunrise-pts}.

In general, coordinate transformations such as~\eqref{eq:shift-blowup} are
also known as resolutions of singularities of algebraic varieties. They can be
computed algorithmically with the help of, for example,
\soft{pySecDec}~\cite{Borowka:2017idc, Heinrich:2021dbf}.

\subsection{Two-loop planar box diagram}
\label{app:dbox-details}
\begin{table}
    \centering
    \begin{tabular}{ccl}
        \toprule
        Chart & Point & Coordinate transformation\brk{s}
        \\\midrule
        \multirow{3}{*}{$U_z$} &
        $\brk{0, 0}$ &
        $\brk{z_1,\> z_2}$
        \\
        &
        $\brk{-t, 0}$ &
        $\brk{
            -t
            + \brk{s - t} \sum_{i = 1}^N \brk{-z_1 z_2 / s}^i
            + z_1 \brk{z_1 z_2}^N
            \> , \>
            z_1 z_2
        }$
        \\
        &
        $\brk{0, -t}$ &
        $\brk{
            z_1 z_2
            \> , \>
            -t
            + \brk{s - t} \sum_{i = 1}^N \brk{-z_1 z_2 / s}^i
            + \brk{z_1 z_2}^{N} z_2
        }$
        \\
        \midrule
        \multirow{1}{*}{$U_x$} &
        \multirow{1}{*}{$\brk{0, 0}$} &
        $\bigbrc{
            \brk{z_1, \tfrac{1}{z_2}}, \>
            \brk{\tfrac{1}{z_2}, \tfrac{1}{z_1 z_2}}, \>
            \brk{
                -s
                - \brk{s - t} \sum_{i = 1}^N \brk{-s \> z_1 z_2}^i
                + z_1 \brk{z_1 z_2}^N
                \> , \>
                \tfrac{1}{z_1 z_2}
            }
        }$
        \\
        \midrule
        \multirow{1}{*}{$U_y$} &
        \multirow{1}{*}{$\brk{0, 0}$} &
        $\bigbrc{
            \brk{\tfrac{1}{z_1 z_2}, \tfrac{1}{z_1}}, \>
            \brk{\tfrac{1}{z_1}, z_2}, \>
            \brk{
                \tfrac{1}{z_1 z_2}
                \> , \>
                -s
                - \brk{s - t} \sum_{i = 1}^N \brk{-s \> z_1 z_2}^i
                + \brk{z_1 z_2}^{N} z_2
            }
        }$
        \\
        \bottomrule
    \end{tabular}
        \caption{
        Intersection points \brk{middle} from the three coordinate
        charts of $\CP^2$ \brk{left} that contribute to the intersection
        numbers between forms~\protect\eqref{eq:forms-massless-sunrise} with
        the massless 2-loop planar box twist~\protect\eqref{eq:twist-dbox}.
        On the right we show the coordinate transformations used to
        compute the residues~\protect\eqref{eq:manyvarsinterX}.
        Some transformations are parametrized by $N \in \Integers_{\ge 0}$.
        We sum over all the contributions of each displayed transformation.
    }
    \label{tab:dbox-pts}
\end{table}

Here we provide further details for \secref{ssec:dbox}, in particular in
\tabref{tab:dbox-pts} we collect the coordinate transformations needed for
evaluation of intersection numbers between forms~\eqref{eq:forms-dbox} with the
twist~\eqref{eq:twist-dbox}.

Consider the point $z = \brk{-t, 0}$ appearing on the second row of \tabref{tab:dbox-pts}.
This point is located at the intersection of the two singular hypersurfaces
$\cS_2 \cap \cS_3$, i.e. satisfies $\cB_2 = \cB_3 = 0$ \brk{see
eq.~\eqref{eq:B-dbox}}.
We look for a pair of local coordinates that roughly looks like
$\brk{z_1, z_2} \mapsto \brk{\cB_3\brk{z_1, z_2}, \cB_2\brk{z_1, z_2}}$\ .
Since $\cB_2\brk{z_1, z_2} \defas z_2$ we may keep $z_2$ as one of our new
coordinates.
To derive the transformation rule for $z_1$ solve:
\begin{align}
    \cB_3\brk{z_1, z_2} \defas
    s t + s \brk{z_1 + z_2} + z_1 z_2 = 0
    \>
    \Longrightarrow
    \>
    z_1 &= -t + \brk{s - t} \> \frac{-z_2 / s}{1 + z_2 / s}
    \ .
\end{align}
and expand this solution near $z_2 = 0$ producing:
\begin{align}
    z_1
    =
    -t + \brk{s - t} \bigbrk{
        -z_2 / s
        + \brk{-z_2 / s}^2
        + \ldots
        + \brk{-z_2 / s}^N
        + \ldots
    }
    \ .
\end{align}
Truncating this expansion up to the $N\textsuperscript{th}$ order and an
additional change $\brk{z_1, z_2} \mapsto \brk{z_1, z_1 z_2}$ reproduces
\brk{up to the $z_1 \brk{z_1 z_2}^N$ term} the coordinate transformation shown
on the second row of~\tabref{tab:dbox-pts}.

In these new coordinates, the twist becomes:
\begin{align}
    \twist =
    z_1^{\gamma_2 + \brk{N + 1} \gamma_3}
    z_2^{\gamma_2 + N \gamma_3}
    \brk{
        -t
        + \ldots
        + z_1 \brk{z_1 z_2}^N
    }^{\gamma_1}
    \bigbrk{
        s
        -\brk{s - t} s^{-N} \> z_2
        + z_1 z_2
    }^{\gamma_3}
    \ ,
\end{align}
where each polynomial factor now has a constant term, so that the series expansion
around $z_1 = z_2 = 0$ is well-defined.
A generic monomial 2-form~\eqref{eq:forms-dbox} transforms as follows:
\begin{align}
    \jj\supbbf{2}_{\bigcdot}
    \defas
    z_1^{n_1} z_2^{n_2}
    \>
    \dd z_1 \wedge \dd z_2
    \mapsto
    z_1^{N + 1 + n_2} z_2^{N + n_2}
    \>
    \brk{
        -t
        + \ldots
        + z_1 \brk{z_1 z_2}^N
    }^{n_1}
    \>
    \dd z_1 \wedge \dd z_2
    \ .
\end{align}
The pole order of the form shifts by $N$ in both variables, which
implies that for large enough $N$ the non-vanishing
condition~\eqref{eq:non-vanishing} will
no longer be satisfied and the corresponding transformation rule will not
contribute to the intersection number.
Hence we conclude, that even though the~\tabref{tab:dbox-pts} collects an
infinite number of coordinate transformations, only a finite number of them
contributes to a given intersection number.

\subsection{Series expansion of the integral formula}
\label{app:closedformint}

Here we derive the expression of a given residue appearing in the main
formula~\eqref{eq:manyvarsinterX} that utilises the series expansion method
presented in~\secref{sec:solution}.
Throughout this section we assume that the pole of the residue is located at
the origin $z = 0$.

\begin{figure}
    \centering
    \subfloat[$\supp\brk{\hat{\jj}_{\bigcdot}}$]{
        \includegraphicsbox[scale=1]{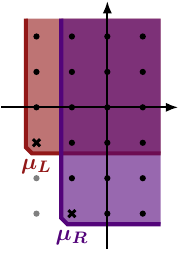}
        \label{fig:cone-lr}
    }
    \subfloat[$\rho_0$]{
        \includegraphicsbox[scale=1]{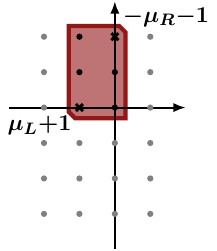}
        \label{fig:cone-rho0}
    }
    \subfloat[$\rho_1$]{
        \includegraphicsbox[scale=1]{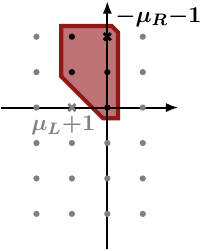}
        \label{fig:cone-rho1}
    }
    \subfloat[$\rho_2$]{
        \includegraphicsbox[scale=1]{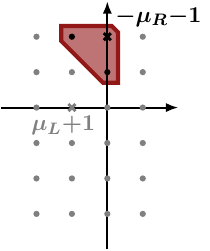}
        \label{fig:cone-rho2}
    }
    \caption{
        Supports of the $\hat{\jj}_L$ and $\hat{\jj}_R$
        expansions~\protect\eqref{eq:forms-expansion} in
        {\protect\subref{fig:cone-lr}},
        and the corresponding summation domains for
        eqs.~\protect\eqref{eq:residue-rho0, eq:residue-rhom}
        in
        \brk{%
            \protect\subref*{fig:cone-rho0},
            \protect\subref*{fig:cone-rho1},
            \protect\subref*{fig:cone-rho2}%
        }.
    }
\end{figure}

\begin{figure}
    \centering
    \subfloat[$\supp\brk{\twist}$]{
        \includegraphicsbox[scale=1]{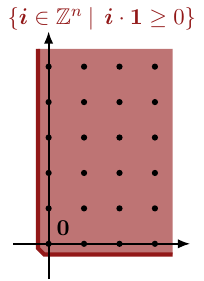}
        \label{fig:cone-u0}
    }
    \subfloat[$\supp\brk{\tilde{\twist}}$]{
        \includegraphicsbox[scale=1]{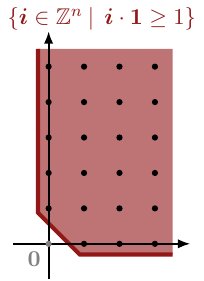}
        \label{fig:cone-u1}
    }
    \subfloat[$\supp\brk{\tilde{\twist}^2}$]{
        \includegraphicsbox[scale=1]{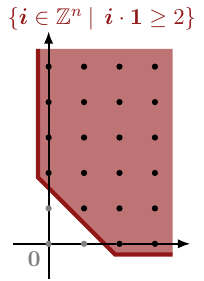}
        \label{fig:cone-u2}
    }
    \subfloat[for eq.~\protect\eqref{eq:residue-rhom}]{
        \includegraphicsbox[scale=1]{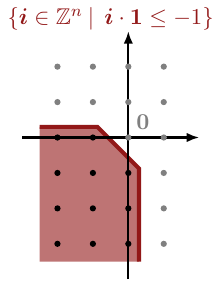}
        \label{fig:cone-u-1}
    }
    \caption{
        Supports of the expansions~\protect\eqref{eq:twist-expansion}
        in \brk{%
            \protect\subref*{fig:cone-u0},
            \protect\subref*{fig:cone-u1},
            \protect\subref*{fig:cone-u2}%
        },
        where we implicitely assume an extra $\vv{i} \ge \vv{0}$ condition.
        In \brk{\protect\subref*{fig:cone-u-1}} we illustrate the notation of
        eq.~\protect\eqref{eq:residue-rhom}
        \brk{an additional $\vv{i} \le \vv{0}$ here is implied as well}.
    }
    \label{fig:cone-u}
\end{figure}

Using the multi-index notation introduced in~\secref{sec:solution}, we write the $z$-expanded cocycles as:
\begin{align}
    \hat{\jj}_L = \sum_{\vv{i} \ge \vv{\mu_{L}}}
    \jj_{L, \vv{i}} \>
    z^\vv{i}
    \ ,
    \quad
    \hat{\jj}_R = \sum_{\vv{i} \ge \vv{\mu_{R}}}
    \jj_{R, \vv{i}} \>
    z^\vv{i}
    \ .
    \label{eq:forms-expansion}
\end{align}
Examples of the domains of summation appearing here are shown
in~\figref{fig:cone-lr}. In the following we will also refer to such domains
of $z$-expansions as supports\footnote{
    Recall that the support of a Laurent series $f = \sum_{\vv{i} \in
    \Integers^n} = f_\vv{i} \> z^\vv{i}$ is the set
    $\supp\brk{f} \defas \brc{\vv{i} \in \Integers^n \>|\> f_\vv{i} \neq 0}$\ .
}.
Expansion of the twist $\twist$ takes the following form:
\begin{align}
    \twist\brk{z} =
    u_0 \> z^\vv{\gamma} \> \bigbrk{1 + \sum_{\vv{i} \cdot \vv{1} \ge 1} u_\vv{i} \>
    z^\vv{i}}
    \equiv
    u_0 \> z^\vv{\gamma} \> \bigbrk{1 + \tilde{\twist}\brk{z}}
    \ ,
    \quad
    \tilde{\twist}\brk{z} \defas \sum_{\vv{i} \cdot \vv{1} \ge 1} u_\vv{i} \> z^\vv{i}
    \ ,
    \label{eq:twist-expansion}
\end{align}
where $u_0$ is a $z$-independent constant, the vector $\vv{\gamma}$ is a
linear combination of exponents appearing in~\eqref{eq:B-def},
$\tilde{\twist}$ collects terms of the expansion that have at
least one power of $z$ inside, and we denote:
\begin{align}
    \vv{i} \cdot \vv{1} = i_1 + \ldots + i_n
    \ .
\end{align}
We show the supports of $\twist$ and $\tilde{\twist}$ in~\figref{fig:cone-u0}
and~\figref{fig:cone-u1} respectively.
In the following we implicitly assume an additional condition $\vv{i} \ge
\vv{j}$ whenever we write $\brk{\vv{i} - \vv{j}} \cdot \vv{1} \ge m$ for some
vector shift $\vv{j}$.

According to the integral formula~\eqref{eq:MMdeqsol}, our goal is to compute
the following residue:
\begin{align}
    \rho \defas \Res_{z = 0} \Bigsbrk{
        \phiR \twist^{-1}
        \int^z
        \phiL \twist
    }
    \label{eq:residue-rho}
\end{align}
The key idea is to expand the inverse of the twist $\twist\brk{z}^{-1}$ in terms
of $\tilde{\twist}\brk{z}$ using the geometric series\footnote{
    For brevity we omit the $z^{-\gamma}$ and $z^{\gamma}$ factors around the
    integral sign that come from the twist~\eqref{eq:twist-expansion}, but they
    can easily be restored via:
    $\int \jj \mapsto z^{-\gamma} \int^z
    z^{\gamma} \jj$
    \brk{with some abuse of notation}.
}:
\begin{align}
    \rho &=
    \underbrace{
        \Res_{z = 0}\Bigsbrk{\phiR \int \phiL}
    }_{\rho_0}
    +
    \sum_{m = 1}^{\brk{
        - \vv{\mu_{L}}
        - \vv{\mu_{R}}
        - \vv{2}}
    \cdot \vv{1}
    } \brk{-1}^m
    \>
    \underbrace{
        \Res_{z = 0}\Bigsbrk{
            \phiR
            \cdot
            \tilde{\twist}^{m - 1}
            \cdot
            \Bigbrk{
                \tilde{\twist} \int \phiL
                -
                \int \tilde{\twist} \phiL
            }
        }
    }_{\rho_m}
    \ ,
    \label{eq:residue-terms}
\end{align}
where in $\rho_m$ we gather the residues with $m$ insertions of
$\tilde{\twist}\brk{z}$.
The summation bounds in the second term follows from the fact that the
$z$-expansion of $\tilde{\twist}\brk{z}^m$ has terms with at least $m$ powers of $z$
inside, as reflected in~\figref{fig:cone-u1}
and~\figref{fig:cone-u2} for $m = 1$ and $m = 2$ respectively.

The residue $\rho_0$ in eq.~\eqref{eq:residue-terms} is straightforward to evaluate:
\begin{align}
    \rho_0
    =
    \sum_{\vv{i} = \vv{\mu_L} + \vv{1}}^{-\vv{\mu_R} - \vv{1}}
    \hat{\jj}_{R, - \vv{i} - \vv{1}} \>
    \hat{\jj}_{L, \vv{i} - \vv{1}} \>
    \frac{1}{\vv{i} + \vv{\gamma}}
    \ ,
    \label{eq:residue-rho0}
\end{align}
where the fraction stands for $1 / \brk{\vv{i} + \vv{\gamma}} \defas 1 / \prod_j
\brk{i_j + \gamma_j}$\ , and the $n$-dimensional summation domain
is shown in~\figref{fig:cone-rho0}.

To compute $\rho_m$ in eq.~\eqref{eq:residue-terms} we separately expand its
three factors, multiply them, and take the residue.
The first factor is given in eq.~\eqref{eq:forms-expansion}.
The second factor is a power of a series, so we may write its
$\vv{i}$\textsuperscript{th} non-zero coefficient
$\tilde{\twist}\brk{z}^m = \sum_\vv{i} \brk{\tilde{\twist}^m}_\vv{i} \> z^\vv{i}$
using the Kronecker symbol $\delta_{\vv{i}, \vv{j}}$ as:
\begin{align}
    \bigbrk{\tilde{\twist}^m}_\vv{i}
    =
    \sum_{{\vv{j_1} \cdot \vv{1}} \ge 1}
    \ldots
    \sum_{{\vv{j_m}} \cdot \vv{1} \ge 1}
    \delta_{\brk{\sum_k \vv{j_k}}, \vv{i}}
    \>
    u_\vv{j_1} \ldots u_\vv{j_m}
    \ ,
    \quad
    \text{for ${{\vv{i} \cdot \vv{1}} \ge m}$ and $\vv{i} \ge \vv{0}$}\ .
    \label{eq:factor-2}
\end{align}
One can come up with various evaluation strategies for this multiple sum using, for
example, the nested parametrization as in eq.~\eqref{eq:rnkgen}, or
the \brk{precomputed} multivariate Bell polynomials~\cite{Withers:2008}.
For the $\vv{i}$\textsuperscript{th} non-zero coefficient of the last factor of
$\rho_m$ we get
\begin{align}
    \Bigbrk{
        \tilde{\twist} \int \phiL
        -
        \int \tilde{\twist} \phiL
    }_\vv{i}
    =
    \sum_{\substack{{\vv{j} \cdot \vv{1}} \ge 1 \\ \vv{j} \le \vv{i} - \vv{\mu_L} - \vv{1}}}
    \twist_\vv{j} \>
    \hat{\jj}_{L, \vv{i} - \vv{j} - \vv{1}} \>
    \Bigbrk{
        \tfrac{1}{\vv{i} - \vv{j} + \vv{\gamma}}
        -
        \tfrac{1}{\vv{i} + \vv{\gamma}}
    }
    \ ,
    \quad
    \text{for ${\brk{\vv{i} - \vv{\mu_L} - \vv{1}} \cdot \vv{1} \ge 1}$}
    \label{eq:factor-3}
\end{align}
and $\vv{i} \ge \vv{\mu_L} + \vv{1}$.
Integrals here are done via the same technique as discussed
in~\secref{sec:solution}.

Finally the sum representation of the $\rho_m$ residue reads:
\begin{align}
    \rho_m
    =
    \sum_{\substack{
        \brk{\vv{i} - \vv{\mu_L} - \vv{1}} \cdot \vv{1} \ge m
        \\
        \vv{i} \le - \vv{\mu_R} - \vv{1}
    }}
    \hat{\jj}_{R, - \vv{i} - \vv{1}}
    \sum_{\substack{
        {\vv{j}} \cdot \vv{1} \ge m - 1
        \\
        \brk{\vv{j} - \vv{i} + \vv{\mu_L} + \vv{1}} \cdot \vv{1} \le -1
    }}
    \brk{\tilde{\twist}^{m - 1}}_\vv{j}
    \Bigbrk{
        \tilde{\twist} \int \phiL
        -
        \int \tilde{\twist} \phiL
    }_{\vv{i} - \vv{j}}
    \ ,
    \label{eq:residue-rhom}
\end{align}
where examples of the outer sum's domain are shown
in figures~\ref{fig:cone-rho1} and \ref{fig:cone-rho2} \brk{see
also~\figref{fig:cone-u-1} for an illustration of the $\le$ notation in the
inner sum}.
Alongside the non-vanishing condition:
\begin{align}
    -\vv{\mu_L} - \vv{\mu_R} - \vv{2} \ge \vv{0}
    \ ,
    \label{eq:non-vanishing}
\end{align}
we obtain the algebraic expression for a given residue:
\begin{align}
    \boxed{
        \Res\brk{\psi \jj_R}
        =
        \sum_{m = 0}^{\brk{
            - \vv{\mu_{L}}
            - \vv{\mu_{R}}
            - \vv{2}}
        \cdot \vv{1}
        } \brk{-1}^m \rho_m
    }
    \label{eq:residue-final}
\end{align}
with $\rho_m$ defined by eqs.~\eqref{eq:residue-rho0}
and~\eqref{eq:residue-rhom, eq:factor-2, eq:factor-3}.

\section{Algebraic solution: proof}
\label{app:proof}

The purpose of this appendix is to show the validity of the algebraic solution given by eq.~\eqref{eq:algebraic} and discuss some features not mentioned in the main text.

\subsection{Deriving the algebraic expression}
\label{app:derivation}

We will consider a more general
case than in~\secref{sec:algebraic}, and only at the end specify to the ``right
rescaling'' framework. Our starting point will be
eqs.~\eqref{eq:manyvarsinterX} and \eqref{eq:mostimportant}:
\begin{align}
    \langle \varphi_L \>|\> \varphi_R \rangle = \sum_{p \in \Poles_\omega} \Res_{z=p} (\psi \varphi_R )
    \label{eq:interxapp}
\end{align}
where $\psi$ is defined through
\begin{align}
    \hat{\nabla}_{\omega_1} \cdots \hat{\nabla}_{\omega_n} \psi = \hat{\varphi}_L \qquad\quad \text{where} \qquad\quad \hat{\nabla}_{\omega_i} := \partial_{z_i} + (\partial_{z_i} \log(u))
    \label{eq:difeqapp}
\end{align}
We want to approach the differential equation locally as a series expansions.
So in the following we will consider a point corresponding to one of the terms
of eq.~\eqref{eq:interxapp} with the assumption that the proper variable
changes has been made so $p = 0$ and in addition that
the series expansions are well defined so no further blowups are needed as also
discussed in~\secref{sec:algebraic}.

Let us write the objects involved\footnote{One might consider why $w$ has
these log terms. Clearly such terms will appear if $u$ has a factor of
$z^\gamma$, and in addition the requirement is that the derivatives of $w$ are
Laurent polynomials, not that $w$ itself is. This is what rules out terms of
the form $z_i \log(z_j)$ as such terms do not have all
derivatives being Laurent. This is also why the branch choice of the logarithms
is irrelevant, since potential factors of $2 \pi i$ will vanish when taking the
derivatives.} as such expansions:
\begin{align}
    \hat{\varphi}_L = \sum_{\boldsymbol{i} = \boldsymbol{\mu}_{\boldsymbol{L}}}^{\boldsymbol{\infty}} \! \hat{\varphi}_{L,\boldsymbol{i}} \, z^{\boldsymbol{i}} \qquad\qquad \hat{\varphi}_R = \sum_{\boldsymbol{i} = \boldsymbol{\mu}_{\boldsymbol{R}}}^{\boldsymbol{\infty}} \! \hat{\varphi}_{R,\boldsymbol{i}} z^{\boldsymbol{i}} \qquad\qquad \psi = \sum_{\boldsymbol{i} = \boldsymbol{\mu}_{\boldsymbol{\psi}}}^{\boldsymbol{\infty}} \! \psi_{\boldsymbol{i}} z^{\boldsymbol{i}}
    \label{eq:phiexpapp}
\end{align}
\begin{align}
    w := \log(u) = \sum_{\boldsymbol{i} = \boldsymbol{\mu}_{\boldsymbol{w}}}^{\boldsymbol{\infty}} \! w_{\boldsymbol{i}} z^{\boldsymbol{i}} \; + \; \sum_{j=1}^n v_j \log(z_j)
    \label{eq:wexpapp}
\end{align}
The exponent of the leading powers $\boldsymbol{\mu}_{\boldsymbol{L}}$, $\boldsymbol{\mu}_{\boldsymbol{R}}$, $\boldsymbol{\mu}_{\boldsymbol{w}}$\footnote{$\boldsymbol{\mu}_{\boldsymbol{w}}$ will be $\boldsymbol{0}$ for all problems where $u = \prod \mathcal{B}^\gamma$ as for the Feynman integrals in Baikov representation discussed in this paper. Yet we keep it general in this appendix since the theory is valid also for other functional forms for $u$. For instance a $u$ containing a factor of $e^{1/z}$ will give $\mu_w = -1$.} are given by the physical problem, while $\boldsymbol{\mu}_{\boldsymbol{\psi}}$ is not since the $\psi$-expression is an ansatz so $\boldsymbol{\mu}_{\boldsymbol{\psi}}$ is just assumed to chosen to be small enough that all manipulations done in the following are valid.

With these expansions we may easily write the residue we are looking for as
\begin{align}
\Res_{z=0}(\psi \varphi_R) = \sum_{\boldsymbol{i} = \boldsymbol{\mu}_{\boldsymbol{\psi}} \!\!}^{-\boldsymbol{\mu}_{\boldsymbol{R}}-\boldsymbol{1} \,} \!\!\! \psi_{\boldsymbol{i}} \, \hat{\varphi}_{R,-\boldsymbol{i}-\boldsymbol{1}}
\label{eq:resexp1}
\end{align}
The challenging part is solving eq.~\eqref{eq:difeqapp} to identify the expressions for the $\psi_{\boldsymbol{i}}$.

\subsubsection{The case $n=2$.}

{\bf Solving the $n$PDE}. We first discuss the case $n=2$, and later we extend it to arbitrary $n$-forms. Solving eq.~\eqref{eq:difeqapp} corresponds to solving $\Delta=0$ with
\begin{align}
\Delta := \hat{\nabla}_{\omega_1} \hat{\nabla}_{\omega_2} \psi - \hat{\varphi}_L
\label{eq:deltexpapp1}
\end{align}
This corresponds to
\begin{align}
\Delta &= (\partial_{1} \partial_{2} \psi) + (\partial_1 \partial_2 w) \psi + (\partial_1 w)(\partial_2 \psi) + (\partial_2 w)(\partial_1 \psi) + (\partial_1 w)(\partial_2 w)\psi - \hat{\varphi}_L
\label{eq:deltexpapp2}
\end{align}
where we have used the obvious notation $\partial_i := \partial_{z_i}$. Inserting the expansions we get
\begin{align}
\Delta &= \;\, \sum_{\boldsymbol{i}=\boldsymbol{\mu}_{\boldsymbol{\psi}}}^{\boldsymbol{\infty}} (i_1 i_2 \psi_{\boldsymbol{i}} - \hat{\varphi}_{L,\boldsymbol{i}-\boldsymbol{1}} ) z_1^{i_1-1} z_2^{i_2-1} \nonumber \\
& \;\; + \sum_{\boldsymbol{i}=\boldsymbol{\mu}_{\boldsymbol{\psi}}}^{\boldsymbol{\infty}} \sum_{\boldsymbol{j}=\boldsymbol{\mu}_{\boldsymbol{w}}}^{\boldsymbol{\infty}} \big( j_1 j_2 w_{\boldsymbol{j}} + i_2 (j_1 w_{\boldsymbol{j}} + \delta_{\boldsymbol{j},\boldsymbol{0}} v_1) + i_1 (j_2 w_{\boldsymbol{j}} + \delta_{\boldsymbol{j},\boldsymbol{0}} v_2) \big) \psi_{\boldsymbol{i}} z_1^{i_1+j_1-1} z_2^{i_2+j_2-1} \nonumber \\
& \;\; + \sum_{\boldsymbol{i}=\boldsymbol{\mu}_{\boldsymbol{\psi}}}^{\boldsymbol{\infty}} \sum_{\boldsymbol{j}=\boldsymbol{\mu}_{\boldsymbol{w}}}^{\boldsymbol{\infty}} \sum_{\boldsymbol{k}=\boldsymbol{\mu}_{\boldsymbol{w}}}^{\boldsymbol{\infty}} \! (j_1 w_{\boldsymbol{j}} + \delta_{\boldsymbol{j},\boldsymbol{0}} v_1) (k_2 w_{\boldsymbol{k}} + \delta_{\boldsymbol{k},\boldsymbol{0}} v_2) \psi_{\boldsymbol{i}} \, z_1^{i_1+j_1+k_1-1} z_2^{i_2+j_2+k_2-1} \label{eq:deltexpapp3}
\end{align}

Changing the summation variables we may rewrite this as
\begin{align}
\Delta &= \!\!\! \sum_{\boldsymbol{s} = \boldsymbol{\mu}_{\boldsymbol{\psi}} + n \boldsymbol{\mu}_{\boldsymbol{w}} \! } \!\!\! \Delta_{\boldsymbol{s}} \, z^{\boldsymbol{s}-\boldsymbol{1}}
\label{eq:deltaexpexp}
\end{align}
where\footnote{In eqs.~\eqref{eq:deltaexpexp} and \eqref{eq:deltasexp} we have reinserted $n$ reflecting the way the corresponding equations will look in the $n$-variate case.}
\begin{align}
\Delta_{\boldsymbol{s}} &:= \sum_{\boldsymbol{t}=\boldsymbol{\mu}_{\boldsymbol{\psi}}}^{\boldsymbol{s}-n\boldsymbol{\mu}_{\boldsymbol{w}}} \!\! R(\boldsymbol{s}{-}\boldsymbol{t},\boldsymbol{t}) \, \psi_{\boldsymbol{t}} \;\; - \;\; \hat{\varphi}_{L,\boldsymbol{s}-\boldsymbol{1}}
\label{eq:deltasexp}
\end{align}
with
\begin{align}
R(\boldsymbol{\alpha}, \boldsymbol{\beta}) &= (\beta_1{+}v_1) (\beta_2{+}v_2) \delta_{\boldsymbol{\alpha},\boldsymbol{0}} + \big( \alpha_1 \alpha_2 + \alpha_1 (\beta_2{+}v_2) + \alpha_2 (\beta_1{+}v_1) \big) w_{\boldsymbol{\alpha}} \nonumber \\
& \;\;\;\, + \sum_{\boldsymbol{j} = \boldsymbol{\mu}_{\boldsymbol{w}}}^{\boldsymbol{\alpha} - \boldsymbol{\mu}_{\boldsymbol{w}}} (\alpha_1 {-} j_1) j_2 \, w_{\boldsymbol{\alpha} - \boldsymbol{j}} w_{\boldsymbol{j}}
\label{eq:rexpapp}
\end{align}
which reduces to the expression given in eq.~\eqref{eq:reks2} in the case where $\boldsymbol{\mu}_{\boldsymbol{w}} = \boldsymbol{0}$. In general this function $R$ may be interpreted as $R(\boldsymbol{\alpha},\boldsymbol{\beta})$ being the coefficient in $\Delta$ of $\psi_{\boldsymbol{\beta}} \, z^{\boldsymbol{\alpha}{+}\boldsymbol{\beta}{-}\boldsymbol{1}}$.

Having $\Delta = 0$ in general requires that each term in the $z$-expansion of eq.~\eqref{eq:deltaexpexp} is zero independently. This means that our task now is to solve $\Delta_{\boldsymbol{s}}=0$ for each value of $\boldsymbol{s}$.

This can be done. If we define
\begin{align}
    \tilde{\boldsymbol{\mu}}_{\boldsymbol{\psi}} := \boldsymbol{\mu}_{\boldsymbol{L}} - n\boldsymbol{\mu}_{\boldsymbol{w}} + \boldsymbol{1}
\end{align}
we find the recursive solution
\begin{align}
    \psi_{\boldsymbol{q}} \; &= \;\qquad\qquad\qquad\qquad\quad 0 \qquad\qquad\qquad\qquad\qquad\qquad \text{if} \;\; \boldsymbol{q} \ngeq \tilde{\boldsymbol{\mu}}_{\boldsymbol{\psi}} \nonumber \\
    \psi_{\boldsymbol{q}} \; &= \;\bigg( \hat{\varphi}_{L,\boldsymbol{q}+n\boldsymbol{\mu}_{\boldsymbol{w}}-\boldsymbol{1}} \; - \!\!\!\!\! \sum_{\boldsymbol{r} :\; \tilde{\boldsymbol{\mu}}_{\boldsymbol{\psi}} \leq \boldsymbol{r} < \boldsymbol{q}} \!\!\!\!\! Q(\boldsymbol{q},\boldsymbol{r}) \, \psi_{\boldsymbol{r}} \bigg) \Big/ Q(\boldsymbol{q},\boldsymbol{q}) \qquad \text{if} \;\; \boldsymbol{q} \geq \tilde{\boldsymbol{\mu}}_{\boldsymbol{\psi}}
    \label{eq:psisolrec}
\end{align}
where
\begin{align}
    Q(\boldsymbol{q},\boldsymbol{r}) := R(\boldsymbol{q}{-}\boldsymbol{r}{+}n \boldsymbol{\mu}_{\boldsymbol{w}} \,,\; \boldsymbol{r} )
\end{align}
and where the exact definitions of $\geq$ and other binary operators applied to
the index vectors are given is~\secref{app:compare}. The quantity $\tilde{\boldsymbol{\mu}}_{\boldsymbol{\psi}}$ may be interpreted as the ``correct'' value for $\boldsymbol{\mu}_{\boldsymbol{\psi}}$, or at least as a value for which $\boldsymbol{\mu}_{\boldsymbol{\psi}} \nleq \tilde{\boldsymbol{\mu}}_{\boldsymbol{\psi}}$ would make it impossible for the above derivation to go through.

The recursive solution given by eq.~\eqref{eq:psisolrec} can be useful on its own to find $\psi$-solutions to use as arguments of the residue function using for instance eq.~\eqref{eq:resexp1}, but let us continue the derivation in order to find a closed expression. \\

\noindent
{\bf Evaluation of the residue.}
The next step is to re-express eq.~\eqref{eq:psisolrec} in a form in which only the coefficients of the individual $\hat{\varphi}_{L,\boldsymbol{i}}$ are recursive. This can be done as
\begin{align}
    \psi_{\boldsymbol{q}} &= \sum_{\boldsymbol{\kappa} = \tilde{\boldsymbol{\mu}}_{\boldsymbol{\psi}}}^{\boldsymbol{q}} \!\! \hat{\varphi}_{L,\boldsymbol{\kappa}{+}n\boldsymbol{\mu}_{\boldsymbol{w}}-\boldsymbol{1}} \, \Psi(\boldsymbol{q},\boldsymbol{q}{-}\boldsymbol{\kappa})
    \label{eq:psiofPsi}
\end{align}
with
\begin{align}
\Psi(\boldsymbol{q},\boldsymbol{\tau}) \; = \left\{ \begin{array}{cc} 0 & \;\; \text{if} \;\; \boldsymbol{\tau} \ngeq \boldsymbol{0} \\ 1/Q(\boldsymbol{q},\boldsymbol{q}) & \;\; \text{if} \;\; \boldsymbol{\tau} = \boldsymbol{0} \\ \frac{-\sum_{\boldsymbol{i}: \, \boldsymbol{0} \leq \boldsymbol{i} < \boldsymbol{\tau}} Q(\boldsymbol{q}{-}\boldsymbol{i}, \boldsymbol{q}{-}\boldsymbol{\tau}) \Psi(\boldsymbol{q},\boldsymbol{i})}{Q(\boldsymbol{q}{-}\boldsymbol{\tau}, \boldsymbol{q}{-}\boldsymbol{\tau})} & \;\; \text{if} \;\; \boldsymbol{\tau} > \boldsymbol{0} \end{array} \right.
\end{align}
or alternatively
\begin{align}
\Psi(\boldsymbol{q},\boldsymbol{\tau}) \; = \left\{ \begin{array}{cc} 0 & \;\; \text{if} \;\; \boldsymbol{\tau} \ngeq \boldsymbol{0} \\ 1/W(\boldsymbol{q},\boldsymbol{0}) & \;\; \text{if} \;\; \boldsymbol{\tau} = \boldsymbol{0} \label{eq:Psirec} \\ \frac{-\sum_{\boldsymbol{i}: \, \boldsymbol{0} \leq \boldsymbol{i} < \boldsymbol{\tau}} W(\boldsymbol{q}{-}\boldsymbol{\tau}, \boldsymbol{\tau}{-}\boldsymbol{i}) \Psi(\boldsymbol{q},\boldsymbol{i})}{W(\boldsymbol{q}{-}\boldsymbol{\tau}, \boldsymbol{0})} & \;\; \text{if} \;\; \boldsymbol{\tau} > \boldsymbol{0} \end{array} \right.
\end{align}
where $W(\boldsymbol{\eta},\boldsymbol{\lambda}) := Q(\boldsymbol{\eta}{+}\boldsymbol{\lambda}, \boldsymbol{\eta})$ or correspondingly
\begin{align}
    W(\boldsymbol{\eta},\boldsymbol{\lambda})
    &= R(\boldsymbol{\lambda}{+}n \boldsymbol{\mu}_{\boldsymbol{w}}, \,\boldsymbol{\eta})
    \label{eq:Wdefapp}
\end{align}
Eq.~\eqref{eq:Psirec} is a recursive expression that only involves the coefficients of the expansion of $w$. The recursion can be solved with the result
\begin{align}
    \Psi(\boldsymbol{q},\boldsymbol{\tau}) &= \sum_{\sigma \in \text{VC}(\boldsymbol{\tau})} \frac{(-1)^{|\sigma|}}{W(\boldsymbol{q},\boldsymbol{0})} \prod_{i=1}^{|\sigma|} \frac{W(\boldsymbol{q}{-}\boldsymbol{\tau}{+}\sum_{j<i} \boldsymbol{\sigma}_{\boldsymbol{j}}, \, \boldsymbol{\sigma}_{\boldsymbol{i}})}{W(\boldsymbol{q}{-}\boldsymbol{\tau}{+}\sum_{j<i} \boldsymbol{\sigma}_{\boldsymbol{j}}, \, \boldsymbol{0})}
    \label{eq:Psisol}
\end{align}
where VC refers to the \textit{vector compositions} defined and discussed
in~\secref{sec:vc}.
By combining eqs.~\eqref{eq:psiofPsi} and \eqref{eq:resexp1}, and changing the summation variables we get

\begin{align}
\text{Res}(\psi \varphi_R)
&= \sum_{\boldsymbol{\tau} = \boldsymbol{0}}^{n \boldsymbol{\mu}_{\boldsymbol{w}} - \boldsymbol{\mu}_{\boldsymbol{L}} - \boldsymbol{\mu}_{\boldsymbol{R}} - \boldsymbol{2} \;} \;\; \sum_{\boldsymbol{h} = \tilde{\boldsymbol{\mu}}_{\boldsymbol{\psi}}}^{-\boldsymbol{\mu}_{\boldsymbol{R}}-\boldsymbol{\tau}-\boldsymbol{1}} \!\!\!\; \Psi(\boldsymbol{h}{+}\boldsymbol{\tau},\boldsymbol{\tau}) \; \hat{\varphi}_{L,\boldsymbol{h}+n \boldsymbol{\mu}_{\boldsymbol{w}}-\boldsymbol{1}} \; \hat{\varphi}_{R,-\boldsymbol{h}-\boldsymbol{\tau}-\boldsymbol{1}}
\label{eq:resexp2}
\end{align}
and finally we may insert the expression for $\Psi$ from eq.~\eqref{eq:Psisol} and we get the most general form of the algebraic expression:
\begin{align}
\boxed{
\text{Res}(\psi \varphi_R)
\; = \!\!\! \sum_{\boldsymbol{\tau} = \boldsymbol{0}}^{n \boldsymbol{\mu}_{\boldsymbol{w}} - \boldsymbol{\mu}_{\boldsymbol{L}} - \boldsymbol{\mu}_{\boldsymbol{R}} - \boldsymbol{2} \;} \!\!\! \sum_{\sigma \in \text{VC}(\boldsymbol{\tau})} \;\; \sum_{\boldsymbol{h} = \boldsymbol{\mu}_{\boldsymbol{L}} - n\boldsymbol{\mu}_{\boldsymbol{w}} + \boldsymbol{1}}^{-\boldsymbol{\mu}_{\boldsymbol{R}}-\boldsymbol{\tau}-\boldsymbol{1}} \!\!\!\!\! Y(\boldsymbol{h},\boldsymbol{\tau},\sigma) \; \hat{\varphi}_{L,\boldsymbol{h}+n \boldsymbol{\mu}_{\boldsymbol{w}}-\boldsymbol{1}} \; \hat{\varphi}_{R,-\boldsymbol{h}-\boldsymbol{\tau}-\boldsymbol{1}}}
\label{eq:algebraicapp}
\end{align}
where
\begin{align}
Y(\boldsymbol{h},\boldsymbol{\tau},\sigma) &:= \frac{(-1)^{|\sigma|}}{W(\boldsymbol{h}{+}\boldsymbol{\tau},\boldsymbol{0})} \prod_{i=1}^{|\sigma|} \frac{W(\boldsymbol{h}{+}\sum_{j<i} \boldsymbol{\sigma}_{\boldsymbol{j}}, \, \boldsymbol{\sigma}_{\boldsymbol{i}})}{W(\boldsymbol{h}{+}\sum_{j<i} \boldsymbol{\sigma}_{\boldsymbol{j}}, \, \boldsymbol{0})}
\label{eq:ydefapp}
\end{align}
We notice that $\boldsymbol{\tau}$ is not really needed as an argument to $Y$ since $\boldsymbol{\tau} = \sum_j \boldsymbol{\sigma}_{\boldsymbol{j}}$.

\subsubsection{Beyond $n = 2$.}
\label{app:Rneq2}

Even though the derivation in the previous section
was done in the case of
$n=2$, it will go through in the same way for different values. In particular
the final result eq.~\eqref{eq:algebraicapp} will look the same for all values
of $n$. The main exceptions are eqs.~\eqref{eq:deltexpapp1,eq:deltexpapp2,eq:deltexpapp3,eq:rexpapp}. Looking briefly at some other cases, for $n=1$ we have
\begin{align}
    \Delta \; = \; \hat{\nabla}_{\omega} \psi - \hat{\varphi}_L \; = \; (\partial_z \psi) + (\partial_z w) \psi - \hat{\varphi}_L
\end{align}
which will correspond to
\begin{align}
    R(\alpha,\beta) = (\beta{+}v) \delta_{\alpha,0} + \alpha w_{\alpha}
\end{align}
as in eq.~\eqref{eq:reks1}. Likewise for $n=3$ we have
\begin{align}
    \Delta &=  \hat{\nabla}_{\omega_1} \hat{\nabla}_{\omega_2} \hat{\nabla}_{\omega_3} \psi - \hat{\varphi}_L \\
    &= (\partial_1 w)(\partial_{23} \psi) + (\partial_2 w)(\partial_{13} \psi) + (\partial_3 w)(\partial_{12} \psi) + (\partial_{12} w)(\partial_{3} \psi) + (\partial_{13} w)(\partial_{2} \psi) + (\partial_{23} w)(\partial_{1} \psi) \nonumber \\
    & + \big( (\partial_1 w) (\partial_{23} w) + (\partial_2 w) (\partial_{13} w) + (\partial_3 w) (\partial_{12} w) \big) \psi + (\partial_1 w) (\partial_2 w) (\partial_3 \psi) + (\partial_1 w) (\partial_3 w) (\partial_2 \psi) \nonumber \\
    & + (\partial_2 w) (\partial_3 w) (\partial_1 \psi) + (\partial_{123} w) \psi + (\partial_1 w) (\partial_2 w) (\partial_3 w) \psi + \partial_{123} \psi - \hat{\varphi}_L
\end{align}
which after a similar derivation yields
\begin{align}
R(\boldsymbol{\alpha}, \boldsymbol{\beta}) &= \; (\beta_1 {+} v_1) (\beta_2 {+} v_2) (\beta_3 {+} v_3) \delta_{\boldsymbol{\alpha},\boldsymbol{0}} \nonumber \\[1mm]
& \;\;\; + \big( (\alpha_1 {+} \beta_1 {+} v_1) (\alpha_2 {+} \beta_2 {+} v_2) (\alpha_3 {+} \beta_3 {+} v_3) - (\beta_1 {+} v_1) (\beta_2 {+} v_2) (\beta_3 {+} v_3) \big) w_{\boldsymbol{\alpha}} \nonumber \\
& \;\;\; + \; \sum_{\boldsymbol{j} = \boldsymbol{\mu}_{\boldsymbol{w}}}^{\boldsymbol{\alpha}{-}\boldsymbol{\mu}_{\boldsymbol{w}}} \big( (\alpha_1{-}j_1) j_2 (j_3{+}\beta_3{+}v_3) + (\alpha_2{-}j_2) j_3 (j_1{+}\beta_1{+}v_1) + (\alpha_3{-}j_3) j_1 (j_2{+}\beta_2{+}v_2) \big) w_{\boldsymbol{\alpha}{-}\boldsymbol{j}} w_{\boldsymbol{j}} \nonumber \\
& \;\;\; + \, \sum_{\boldsymbol{j} = \boldsymbol{\mu}_{\boldsymbol{w}}}^{\boldsymbol{\alpha} - 2 \boldsymbol{\mu}_{\boldsymbol{w}}} \, \sum_{\boldsymbol{l} = \boldsymbol{\mu}_{\boldsymbol{w}}}^{\boldsymbol{\alpha}{-}\boldsymbol{j}{-}\boldsymbol{\mu}_{\boldsymbol{w}}} \, (\alpha_1{-}j_1{-}l_1) \, j_2 \, l_3 \, w_{\boldsymbol{\alpha}{-}\boldsymbol{j}{-}\boldsymbol{l}} \, w_{\boldsymbol{j}} \, w_{\boldsymbol{l}}
\label{eq:reks3app}
\end{align}
which reduces to eq.~\eqref{eq:reks3} in the case of $\boldsymbol{\mu}_{\boldsymbol{w}} = \boldsymbol{0}$.

Going through the derivation for various values of $n$ allows us to realize the pattern and find the general expression for $R$ which is given in the following.

\subsubsection*{General expression for $R$}

For $R$ in the $n$-variate case, we may write
\begin{align}
R(\boldsymbol{\alpha},\boldsymbol{\beta}) = \sum_{k=0}^n R_{n,k}
\label{eq:Rgeneraln}
\end{align}
where $R_{n,k}$ contains the terms with $k$ $w$-factors and correspondingly $k{-}1$ sums. We find it convenient to use distinct expressions for $R_{n,k}$ for $k=0$, $k=1$, and\footnote{We could in principle get the $k=1$ expression as a special case of eq.~\eqref{eq:rnkgen}.} $k \geq 2$.
Defining
\begin{align}
\lambda_i := \beta_i+v_i
\end{align}
we have
\begin{align}
R_{n,0} &= \Big( \prod_{i=1}^n \lambda_i \Big) \delta_{\boldsymbol{\alpha},\boldsymbol{0}}
\end{align}
\begin{align}
R_{n,1} &= \bigg( \Big( \prod_{i=1}^n (\lambda_i{+}\alpha_i) \Big) - \Big( \prod_{i=1}^n \lambda_i \Big) \bigg) w_{\boldsymbol{\alpha}}
\end{align}
\begin{align}
R_{n,k} &= \!\!\! \sum_{\boldsymbol{i}_{\boldsymbol{2}} = \boldsymbol{\mu}_{\boldsymbol{w}}}^{\boldsymbol{\alpha} - (k-1)\boldsymbol{\mu}_{\boldsymbol{w}}} \cdots \!\!\!\!\!\! \sum_{\boldsymbol{i}_{\boldsymbol{j}} = \boldsymbol{\mu}_{\boldsymbol{w}}}^{\boldsymbol{\alpha} - \sum_{q=2}^{j-1} \boldsymbol{i}_{\boldsymbol{q}} - (k+1-j)\boldsymbol{\mu}_{\boldsymbol{w}}} \!\!\!\!\!\! \cdots \sum_{\boldsymbol{i}_{\boldsymbol{k}} = \boldsymbol{\mu}_{\boldsymbol{w}}}^{\boldsymbol{\alpha} {-} \sum_{q=2}^{k-1} \boldsymbol{i}_{\boldsymbol{q}} - \boldsymbol{\mu}_{\boldsymbol{w}}} \nonumber \\
& \quad S_{n,k} \Big( \Big( \boldsymbol{\alpha} {-} \sum_{q=2}^k \boldsymbol{i}_{\boldsymbol{q}} \Big), \boldsymbol{i}_{\boldsymbol{2}}, \ldots, \boldsymbol{i}_{\boldsymbol{k}} , \boldsymbol{\lambda} \Big) \, w_{( \boldsymbol{\alpha} {-} \sum_{q=2}^k \boldsymbol{i}_{\boldsymbol{q}})} w_{\boldsymbol{i}_{\boldsymbol{2}}} \cdots w_{\boldsymbol{i}_{\boldsymbol{k}}}
\label{eq:rnkgen}
\end{align}
where we have defined
\begin{align}
S_{n,k}(\boldsymbol{i}_{\boldsymbol{1}}, \boldsymbol{i}_{\boldsymbol{2}}, \ldots, \boldsymbol{i}_{\boldsymbol{k}} , \boldsymbol{\lambda} ) &:= \!\!\!\! \sum_{s \in \text{KSN}(n,k)} \!\! \bigg( \prod_{l \in \text{CS}(n,s)} \!\!\!\!\!\! \lambda_l \, \bigg) \prod_{y=1}^k \prod_{j \in s_y} i_{y,j}
\end{align}
Here we have $\text{KSN}(n,k)$ being defined as the set of all sets with size $k$ of (non-empty and non-overlapping) subsets of the set of integers from 1 to $n$. An example being
\begin{align}
\text{KSN}(3,2) &= \Big\{ \big\{ \{1\}, \{2\} \big\}, \big\{ \{1\}, \{3\} \big\}, \big\{ \{2\}, \{3\} \big\}, \nonumber \\
& \qquad \big\{ \{1\}, \{2,3\} \big\}, \big\{ \{2\}, \{1,3\} \big\}, \big\{ \{3\}, \{1,2\} \big\} \Big\}
\end{align}
where it should be possible to see the structure reflected in the single-sum term of eq.~\eqref{eq:reks3app}. The other set-function $\text{CS}(n,s)$ gives the counter-set to one of these sets, that is
\begin{align}
\text{CS}(n,s) := \{1,\ldots,n\} / \bigcup_i s_i
\end{align}
with an example being
\begin{align}
\text{CS}(3,\big\{\{1\},\{3\}\big\}) &= \{2\}
\end{align}

We should point out that it is possible to write eq.~\eqref{eq:rnkgen} nicer and more symmetrically at the cost of introducing an extra sum and a delta function, that is\footnote{
    For practical implementations one can use finite upper limits and
    replace
    $\boldsymbol{\infty} \rightarrow \boldsymbol{\alpha} - (k{-}1)
    \boldsymbol{\mu}_{\boldsymbol{w}}$.
}
\begin{align}
R_{n,k} &= \!\!\! \sum_{\boldsymbol{i}_{\boldsymbol{1}} = \boldsymbol{\mu}_{\boldsymbol{w}}}^{\boldsymbol{\infty}} \!\! \cdots \!\! \sum_{\boldsymbol{i}_{\boldsymbol{k}} = \boldsymbol{\mu}_{\boldsymbol{w}}}^{\boldsymbol{\infty}} \delta_{(\sum_j \! \boldsymbol{i}_{\boldsymbol{j}}),\boldsymbol{\alpha}} \, S_{n,k} ( \boldsymbol{i}_{\boldsymbol{1}}, \ldots, \boldsymbol{i}_{\boldsymbol{k}} , \boldsymbol{\lambda} ) \, w_{\boldsymbol{i}_{\boldsymbol{1}}} \cdots w_{\boldsymbol{i}_{\boldsymbol{k}}}
    \label{eq:rnkgendelta}
\end{align}
where evaluating the $\boldsymbol{i}_{\boldsymbol{1}}$-sum would bring back eq.~\eqref{eq:rnkgen}.

\subsection{Algebraic expression in the rescaling frameworks}

In~\secref{sec:rescaling} it was discussed how to simplify the computation of
the intersection number by rescaling the quantities on which it depends,
locally at each intersection point. Two different cases were presented, right rescaling and left rescaling, which we will discuss below.

\subsubsection*{Right rescaling}

Right rescaling is defined by the rescalings and renamings given in eq.~\eqref{eq:rightrescalingdef}
\begin{align}
\hat{\varphi}_L \rightarrow \hat{\phi} = \hat{\varphi}_L \hat{\varphi}_R \,,\qquad
\hat{\varphi}_R \rightarrow 1 \,,\qquad
u \rightarrow u_R = u/\hat{\varphi}_R \,,\qquad
\psi \rightarrow f \,.
\end{align}
Eq.~\eqref{eq:algebraicapp} may then be used for these rescaled variables, with $\hat{\varphi}_{L,\boldsymbol{i}}$, $\hat{\varphi}_{R,\boldsymbol{i}}$, $w_{\boldsymbol{i}}$, and $v_{j}$ interpreted as the coefficients in the expansions of the rescaled variables $\hat{\phi}$, $1$, and $\log(u_R)$, and $\boldsymbol{\mu}_{\boldsymbol{L}}$, $\boldsymbol{\mu}_{\boldsymbol{R}}$, $\boldsymbol{\mu}_{\boldsymbol{w}}$ as the corresponding powers of the leading coefficients. But the expansion of $1$ is of course trivial, corresponding to $\hat{\varphi}_{R,\boldsymbol{i}} = \delta_{\boldsymbol{i},\boldsymbol{0}}$ and $\boldsymbol{\mu}_{\boldsymbol{R}}=0$. This allows us to do the $h$-sum of eq.~\eqref{eq:algebraicapp} giving
\begin{align}
\text{Res}(f)
\; = \!\!\!\!\! \sum_{\boldsymbol{\tau} = \boldsymbol{0}}^{n \boldsymbol{\mu}_{\boldsymbol{w}} - \boldsymbol{\mu}_{\boldsymbol{L}} - \boldsymbol{2} } \!\!\!\! \sum_{\sigma \in \text{VC}(\boldsymbol{\tau})} \!\!\!\!\! Y(-\boldsymbol{\tau}{-}\boldsymbol{1},\boldsymbol{\tau},\sigma) \;\, \hat{\phi}_{n \boldsymbol{\mu}_{\boldsymbol{w}} - \boldsymbol{\tau} - \boldsymbol{2}}
\end{align}
and inserting the expression for $Y$ the residue becomes
\begin{align}
\text{Res}(f)
\; = \!\!\!\!\! \sum_{\boldsymbol{\tau} = \boldsymbol{0}}^{n \boldsymbol{\mu}_{\boldsymbol{w}} - \boldsymbol{\mu}_{\boldsymbol{L}} - \boldsymbol{2} } \!\!\!\! \sum_{\sigma \in \text{VC}(\boldsymbol{\tau})} \frac{(-1)^{|\sigma|}}{U(\boldsymbol{0},\boldsymbol{0})} \prod_{i=1}^{|\sigma|} \frac{U(\boldsymbol{\sigma}_{\boldsymbol{i}}, \, \boldsymbol{\tau}{-}\sum_{j<i} \boldsymbol{\sigma}_{\boldsymbol{j}})}{U(\boldsymbol{0}, \, \boldsymbol{\tau}{-}\sum_{j<i} \boldsymbol{\sigma}_{\boldsymbol{j}})} \;\, \hat{\phi}_{n \boldsymbol{\mu}_{\boldsymbol{w}}{-}\boldsymbol{\tau}{-}\boldsymbol{2}}
\label{eq:cfrightapp}
\end{align}
where
\begin{align}
    U(\boldsymbol{\alpha},\boldsymbol{\beta}) &= W(-\boldsymbol{\beta}{-}\boldsymbol{1},\, \boldsymbol{\alpha}) \nonumber \\
    &= R(\boldsymbol{\alpha} {+} n \boldsymbol{\mu}_{\boldsymbol{w}} ,\, -\boldsymbol{\beta}{-}\boldsymbol{1})
\end{align}

Inserting $\boldsymbol{\mu}_{\boldsymbol{w}} = \boldsymbol{0}$, eq.~\eqref{eq:cfrightapp} reduces to the algebraic expression discussed in the main text in eq.~\eqref{eq:algebraic}.

\subsubsection*{Left rescaling}

One may go through the same steps for the left rescaling framework, which is defined in terms of the renamings and rescalings defined in eq.~\eqref{eq:leftrescalingdef}
\begin{align}
\hat{\varphi}_L \rightarrow 1 \,,\qquad
\hat{\varphi}_R \rightarrow \hat{\phi} = \hat{\varphi}_L \hat{\varphi}_R \,,\qquad
u \rightarrow u_L = u \hat{\varphi}_L \,,\qquad
\psi \rightarrow g \,.
\end{align}
Inserting this in eq.~\eqref{eq:algebraicapp}, we get
\begin{align}
\text{Res}(g \hat{\varphi}_L \hat{\varphi}_R)
\; = \!\!\! \sum_{\boldsymbol{\tau} = \boldsymbol{0}}^{n \boldsymbol{\mu}_{\boldsymbol{w}} - \boldsymbol{\mu}_{\boldsymbol{R}} - \boldsymbol{2} \;} \!\!\!\! \sum_{\sigma \in \text{VC}(\boldsymbol{\tau})} \frac{(-1)^{|\sigma|}}{\tilde{U}(\boldsymbol{0},\boldsymbol{\tau})} \prod_{i=1}^{|\sigma|} \frac{\tilde{U}(\boldsymbol{\sigma}_{\boldsymbol{i}}, \, \sum_{j<i} \boldsymbol{\sigma}_{\boldsymbol{j}})}{\tilde{U}(\boldsymbol{0},\, \sum_{j<i} \boldsymbol{\sigma}_{\boldsymbol{j}})} \;\, \hat{\phi}_{n \boldsymbol{\mu}_{\boldsymbol{w}} {-}\boldsymbol{\tau}{-}\boldsymbol{2}}
\label{eq:cfleftapp}
\end{align}
with
\begin{align}
    \tilde{U}(\boldsymbol{\alpha},\boldsymbol{\beta}) &= W(\boldsymbol{\beta}{+}\boldsymbol{1}{-}n\boldsymbol{\mu}_{\boldsymbol{w}} ,\, \boldsymbol{\alpha}) \nonumber \\
    &= R(\boldsymbol{\alpha}{+}n \boldsymbol{\mu}_{\boldsymbol{w}} ,\, \boldsymbol{\beta}{+}\boldsymbol{1}{-}n\boldsymbol{\mu}_{\boldsymbol{w}})
\end{align}
Eq.~\eqref{eq:cfleftapp} is the algebraic expression in the left rescaling framework.

\subsection{Additional details}

\subsubsection*{Expansion maxima}

In eqs.~\eqref{eq:phiexpapp,eq:wexpapp} we wrote the expansions of
$\hat{\varphi}_L$, $\hat{\varphi}_R$, and $w$ as formal expansions from the
corresponding $\boldsymbol{\mu}$ up to $\boldsymbol{\infty}$. Yet we do not
need infinitely many terms in the series, there is a maximal value above which
the terms do not contribute to the result of eq.~\eqref{eq:algebraicapp}.
Considering eq.~\eqref{eq:algebraicapp} as well as the general $n$ expression
for $R$ given in and below eq.~\eqref{eq:Rgeneraln} we may deduce what these maximal values are. Using the notation $\boldsymbol{M}$ in parallel with that used for the minimum values $\boldsymbol{\mu}$ we may write an expansion only containing the terms that are actually needed:
\begin{align}
    \hat{\varphi}_L = \sum_{\boldsymbol{i} = \boldsymbol{\mu}_{\boldsymbol{L}}}^{\boldsymbol{M}_{\boldsymbol{L}}} \! \hat{\varphi}_{L,\boldsymbol{i}} z^{\boldsymbol{i}} \qquad\quad \hat{\varphi}_R = \sum_{\boldsymbol{i} = \boldsymbol{\mu}_{\boldsymbol{R}}}^{\boldsymbol{M}_{\boldsymbol{R}}} \! \hat{\varphi}_{R,\boldsymbol{i}} z^{\boldsymbol{i}} \qquad\quad w = \sum_{\boldsymbol{i} = \boldsymbol{\mu}_{\boldsymbol{w}}}^{\boldsymbol{M}_{\boldsymbol{w}}\!} \! w_{\boldsymbol{i}} z^{\boldsymbol{i}} \, + \, \sum_{j=1}^n v_j \log(z_j)
\end{align}
where we have
\begin{align}
\boldsymbol{M}_{\boldsymbol{L}} = n \boldsymbol{\mu}_{\boldsymbol{w}} {-} \boldsymbol{\mu}_{\boldsymbol{R}} {-} \boldsymbol{2} \,, \qquad\quad
\boldsymbol{M}_{\boldsymbol{R}} = n \boldsymbol{\mu}_{\boldsymbol{w}} {-} \boldsymbol{\mu}_{\boldsymbol{L}} {-} \boldsymbol{2} \,, \qquad\quad
\boldsymbol{M}_{\boldsymbol{w}} = (n{+}1) \boldsymbol{\mu}_{\boldsymbol{w}} {-} \boldsymbol{\mu}_{\boldsymbol{L}} {-} \boldsymbol{\mu}_{\boldsymbol{R}} {-} \boldsymbol{2} \,.
\end{align}
This fact is useful for implementation purposes, since it tells which coefficients to extract before applying the algebraic expression. Applying one of the rescaled versions of the algebraic expression does not change these limits if expressed in terms of the original $\varphi_L$ and $\varphi_R$.

\subsubsection*{The univariate case}

In the univariate case there are two simplifications taking place which deserve
discussion. The first is that the vector compositions VC reduce to standard compositions as discussed in~\secref{sec:vc}. The second is that only one $\omega$ is present which makes it desirable to express the result in terms of expansions of that object, replacing the $w$-expansion of eq.~\eqref{eq:wexpapp} with its separately treated log-terms. That is
\begin{align}
    \omega = \sum_{i = \mu_{\omega}}^{\infty} \! \omega_i \, z^i
\end{align}
with the relation between the two expansions being $\omega_i = (i{+}1) w_{i{+}1}$ except for $\omega_{-1} = v_1$. We also have $\mu_{\omega} = \mu_{w}{-}1$.
Inserting this in eq.~\eqref{eq:algebraicapp} we get
\begin{align}
\text{Res}(\psi \varphi_R)
\; = \!\!\!\! \sum_{\tau = 0}^{\mu_{\omega} - \mu_{\phi} - \mu_{\xi} - 1 \;} \!\!\!\! \sum_{\sigma \in \text{comp}(\tau)} \;\; \sum_{h = \mu_{\phi} - \mu_{\omega}}^{-\mu_{\xi}-\tau-1} \!\!\!\!\! Y(h,\tau,\sigma) \; \hat{\varphi}_{L,h+\mu_{\omega}} \; \hat{\varphi}_{R,-h-\tau-1}
\end{align}
where
\begin{align}
Y(h,\tau,\sigma) &= \frac{(-1)^{|\sigma|}}{W(h{+}\tau,0)} \prod_{i=1}^{|\sigma|} \frac{W(h{+}\sum_{j<i} \sigma_{j}, \, \sigma_{i})}{W(h{+}\sum_{j<i} \sigma_{j}, \, 0)}
\end{align}
as in eq.~\eqref{eq:ydefapp} but with
\begin{align}
W(\eta,\lambda) = \eta \, \delta_{\lambda+\mu_{\omega},-1} + \omega_{\lambda+\mu_{\omega}}
\end{align}

\subsubsection*{Further discussion}

It is worthwhile to look at the special case in which everything (e.g. $\varphi_L$, $\varphi_R$, and all $\omega_i$) has simple poles only. This corresponds to $\boldsymbol{\mu}_{\boldsymbol{L}} = \boldsymbol{\mu}_{\boldsymbol{R}} = \boldsymbol{-1}$ and $\boldsymbol{\mu}_{\boldsymbol{w}} = \boldsymbol{0}$. Inserting this in eq.~\eqref{eq:algebraicapp} we get that each of the sums contain one term only, leaving
\begin{align}
    \text{Res}(\psi \varphi_R) &= \frac{\hat{\varphi}_{L,\boldsymbol{-1}} \, \hat{\varphi}_{R,\boldsymbol{-1}}}{\prod_{i=1}^n v_i}
\end{align}
Combining this with eq.~\eqref{eq:manyvarsinterX} we get an expression for the intersection number similar to an expression from refs.~\cite{matsumoto1998, Mizera:2017rqa}.

Additionally it is worth noticing that that the intersection number is
\textit{linear} in both $\varphi_L$ and $\varphi_R$ which may be seen from the
integral definition in eq.~\eqref{eq:interx-def}. That property is easy to see
in the algebraic expression as it is given by eq.~\eqref{eq:algebraicapp}. But
in the rescaling framework linearity becomes hidden.
For instance looking at the solution in the right-rescaling framework as it is
given by eq.~\eqref{eq:algebraic}, linearity in $\jj_L$ is obvious. Linearity
in $\jj_R$, however,
becomes obscured as the $\jj_{R, \boldsymbol{i}}$-terms mix
with the $w_{\boldsymbol{i}}$-factors, which enter the expressions in a very non-linear fashion.

\subsection{Comparing index vectors}
\label{app:compare}

\begin{figure}
    \centering
    \subfloat[$\; \boldsymbol{a} \ge \boldsymbol{b}$]{
        \includegraphicsbox[scale=1]{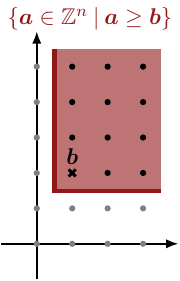}
    }
    \subfloat[$\; \boldsymbol{a} > \boldsymbol{b}$]{
        \includegraphicsbox[scale=1]{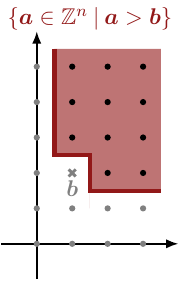}
    }
    \subfloat[$\; \boldsymbol{a} \leq \boldsymbol{b}$]{
        \includegraphicsbox[scale=1]{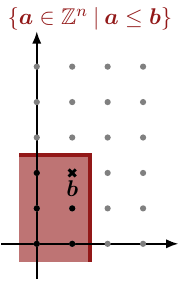}
    }
    \subfloat[$\; \boldsymbol{a} \nleq \boldsymbol{b}$]{
        \includegraphicsbox[scale=1]{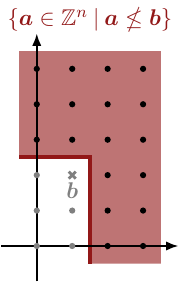}
    }
    \caption{
        Illustration of some of the binary operators defined in~\tabref{tab:compareindices}. The shaded areas cover the integer values for which the illustrated relation is true.
    }
    \label{fig:cone}
\end{figure}

\begin{table}[h!]
  \begin{center}
    \begin{tabular}{|c|c|}
      \hline
      $\boldsymbol{a} = \boldsymbol{b}$ & $\forall i$: $a_i = b_i$ \\ \hline
      $\boldsymbol{a} \leq \boldsymbol{b}$ & $\forall i$: $a_i \leq b_i$ \\ \hline
      $\boldsymbol{a} \geq \boldsymbol{b}$ & $\forall i$: $a_i \geq b_i$ \\ \hline
      $\boldsymbol{a} < \boldsymbol{b}$ & $\boldsymbol{a} \leq \boldsymbol{b}$ and $\boldsymbol{a} \neq \boldsymbol{b}$ \\ \hline
    \end{tabular} \,
    \begin{tabular}{|c|c|}
      \hline
      $\boldsymbol{a} \neq \boldsymbol{b}$ & $\text{not}(\boldsymbol{a} = \boldsymbol{b})$ \\ \hline
      $\boldsymbol{a} \nleq \boldsymbol{b}$ & $\text{not}(\boldsymbol{a} \leq \boldsymbol{b})$ \\ \hline
      $\boldsymbol{a} \ngeq \boldsymbol{b}$ & $\text{not}(\boldsymbol{a} \geq \boldsymbol{b})$ \\ \hline
      $\boldsymbol{a} > \boldsymbol{b}$ & $\boldsymbol{a} \geq \boldsymbol{b}$ and $\boldsymbol{a} \neq \boldsymbol{b}$ \\ \hline
    \end{tabular}
    \caption{The definitions needed to compare index vectors. \label{tab:compareindices}}
  \end{center}
\end{table}

In this appendix and in~\secref{sec:algebraic} we used the notation for
the $n$-variate formula eq.~\eqref{eq:algebraic} where the sums and products
are parameterized by index vectors written in bold, of length $n$ and
containing integers. Such vectors can be compared using the so-called {\it product
order}, let us briefly review how it is done.
If we let $\boldsymbol{a}$ and $\boldsymbol{b}$ be two such index vectors of
length $n$, the relations we use in this paper are defined as
in~\tabref{tab:compareindices}, and some are illustrated in~\figref{fig:cone}.  We see that $\nleq$ and $>$ are not equivalent, for
instance it is true that $\binom{1}{2} \nleq \binom{2}{0}$ while it is false
that $\binom{1}{2} > \binom{2}{0}$. We notice that the index vectors form a
\textit{partially ordered set} under the $\leq$ operator.

\bibliographystyle{JHEP}
\bibliography{biblio}

\end{document}